\documentclass{article}

\usepackage{arxiv}

\usepackage[utf8]{inputenc} 
\usepackage[T1]{fontenc}    
\usepackage{hyperref}       
\usepackage{url}            
\usepackage{booktabs}       
\usepackage{amsfonts}       
\usepackage{nicefrac}       
\usepackage{microtype}      
\usepackage{lipsum}		
\usepackage{graphicx}
\usepackage{natbib}
\usepackage{doi}
\usepackage{tabularx}

\title{MOTENS: A Pedagogical Design Model for Serious Cyber Games}

\author{ \href{https://orcid.org/0000-0001-5108-7116}{\includegraphics[scale=0.06]{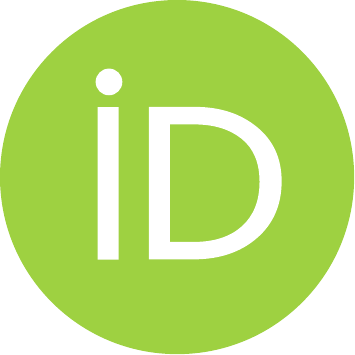}\hspace{1mm}Stephen Hart}\thanks{Corresponding author.} \\
	Electronics and Computer Science\\
	University of Southampton\\ 
	University Road\\ 
	Southampton\\ 
	United Kingdom\\ 
	SO17 1BJ\\
	\texttt{stephen.hart@soton.ac.uk} \\
	\And
	\href{https://orcid.org/0000-0003-3470-7226}{\includegraphics[scale=0.06]{orcid.pdf}\hspace{1mm}Basel Halak} \\
		Electronics and Computer Science\\
	University of Southampton\\ 
	University Road\\ 
	Southampton\\ 
	United Kingdom\\ 
	SO17 1BJ\\
	\texttt{Basel.Halak@soton.ac.uk} \\
	\AND
	\href{https://orcid.org/0000-0002-6432-1482}{\includegraphics[scale=0.06]{orcid.pdf}\hspace{1mm}Vladimiro Sassone} \\
	Electronics and Computer Science\\
	University of Southampton\\ 
	University Road\\ 
	Southampton\\ 
	United Kingdom\\ 
	SO17 1BJ\\
	\texttt{vsassone@soton.ac.uk} \\
	
}

\renewcommand{\headeright}{}
\renewcommand{\undertitle}{}
\renewcommand{\shorttitle}{MOTENS: A Pedagogical Design Model}

\hypersetup{
pdftitle={MOTENS: A Pedagogical Design Model for Serious Cyber Games},
pdfsubject={q-bio.NC, q-bio.QM},
pdfauthor={Stephen Hart, Basel Halak, Vladimiro Sassone},
pdfkeywords={Serious gaming, Cyber security education, Gamification, Design, Training effectiveness},
}

\begin{document}
\maketitle

\begin{abstract}
In the last few years, serious games have become popular, with a consensus of the benefits for teaching cyber security awareness and education. However, there is still a lack of pedagogical driven methodologies and tools to support serious games design to ensure they achieve the learning objectives. This paper proposes MOTENS, a pedagogical model, to design serious cyber games based on the gaps we identified in the current games design models and the lessons learnt from creating a serious tabletop game called Riskio, designed to teach cyber security awareness and education. The MOTENS model has six high-level components. Five components are linked to the games/design mechanics, and one component, `Theory', that supports the design's cognitive principles, including players' motivation. The model is used to design serious cyber games and goes through five stages, from identifying and segmenting target players, steps to creating game mechanics linked to pedagogy instruction and then to testing to create a serious game that is designed to achieve the games learning objectives. 
\end{abstract}


\section{Introduction}
\label{sec:introduction}
In the last few years, serious games have become popular based on gamification principles, which is about applying game mechanics to non-gaming activities, for example, training to make the activity more engaging \citep{routledge2016games}. Serious games use these techniques to provide a fun, enjoyable educational environment where the game participants learn by playing the game.  
Using serious games has become a technique used in various contexts to motivate people to engage in mainly targeted behaviours \citep{landers2014developing}. Serious games as an approach that can complement instruction-led or computer-based cyber security education training~\citep{hart2020riskio}. For example, there have been many media reports of attacks against large and small organisations, causing financial losses and reputational damage as a response to increasing cyber attacks~\citep{buil2021cybercrime}. Organisations invest in professional training courses for their employees to raise awareness of cyber attacks and related defences. However, these traditional courses have failed in effectively educating employees, as testified by the increasing number of successful cyber attacks exploiting human factors~\citep{angafor2020bridging}. The traditional approach to education does not present problems to students but presents contents to resolve problems. In 1994 research survey with academic staff (n = 65) found that only 27\% gave high rating in considering the `learner', where content scored 65\% \citep{seng2004students}. It is fundamental for organisations to ensure that all employees are educated on cyber security concepts and aware of the risks posed by even the simplest cyber attacks, e.g. phishing emails. Organisations have used tailored training and awareness programs to improve the resilience to cyber attacks against their employees'. Serious cyber games can only do this if designed to create an adaptive learning experience such as role-playing, simulations, and self-paced or team-based exercises~\citep{greitzer2007cognitive} and move to problem-based learning. The main drive to problem-based is an emphasis on solving real-world problems~\citep{seng2004students}. 
Threat actors are continuously improving their cyber weapons to timely and effectively exploit vulnerabilities~\citep{hart2020riskio}. Problem-based learning has been criticised for its emphasis on Bloom's \citep{bloom1956taxonomy} higher order thinking skills at the expense of lower-order skills and knowledge acquisition \citep{hung2008problem}. Serious cyber games can provide a good environment for self-learning. However, they must be designed using cognitive principles for effective learning \citep{greitzer2007cognitive}. 

There is consensus on the benefits to the potential use of serious cyber games for awareness and education \citep{arnab2015mapping}. However, there is still a lack of pedagogical driven methodologies and tools to support serious games \citep{arnab2015mapping}. Some of the critical issues with pedagogical models: they do not link game mechanics to the learning objectives; high-level model and will not assist in the selection of game mechanics to achieve serious game objectives; or are mainly assessed in terms of the quality of their content, not in terms of their intention-based design \citep{mitgutsch2012purposeful}. For example, the GOM Model~\citep{amory2007game} describes a relationship between the pedagogical dimensions of learning and game elements but has no concept for the design principles of education games \citep{amory2003educational}. The SGDAF model~\citep{mitgutsch2012purposeful} does not link learning mechanics to game mechanics and LM-GM~\citep{arnab2015mapping}, which proved to be effective at mapping the pedagogical principles for serious cyber games. Still, it is complex to understand and proven to take time to learn.  

This paper proposes a new pedagogical model to design serious cyber games for awareness and education called MOTENS. This was created based on the gaps identified in the current games and experience of lessons learnt from the creation of Riskio \citep{hart2020riskio}. The model was designed to assist design serious cyber games for education and awareness rather than other categories of serious games, for example, secure software development.
\newline 
\newline
The rest of this paper is organised as follows. \autoref{sec:cognitive_principles} outlines the cognitive principles that support gamification. \autoref{sec:motens_model} proposes a new pedagogical model for serious cyber games design based on gaps identified in \autoref{sec:cognitive_principles}. \autoref{sec:case_study} is an  illustrative case study to test the efficacy of the MOTENS Model. \autoref{sec:case_study2} is a second case study is comparative case study to validate the proposed model against another model. Finally, \autoref{sec:summary} concludes the paper. 
\section{Cognitive Principles}
\label{sec:cognitive_principles}
In this section we present the cognitive principles that supports gamification and learning theory and then identify gaps in current models and use game we created called Riskio \citep{hart2020riskio} to map to serious games design models. 

\subsection{Essentials for learning}
Robsen et al proposed a psychology model behind the promise of gamification, which includes three principles for creating gamification experiences: mechanics, dynamics, and emotions (MDE)~\citep{robson2015all}. Mechanics (i.e., the goals, rules, and rewards), dynamics (i.e., how players interact with the game mechanics), and emotions (i.e., how players feel toward the game playing experience). The game mechanics requirements can change based on games players, for example, in the Riskio study students wanted more game-like play in the game mechanics and wanted the attack suit to be random. In contrast, employees were happy with game mechanics, where they were allowed to select the attack suit~\citep{hart2020riskio}. The dynamics are about how the player progresses in the game mechanics and how they feel about the interaction. In-game design, ‘aesthetics’ describes the desirable emotional response~\citep{robson2015all}. The graphic design, quality of illustration, can significantly affect the initial emotional reception of the game. The players' emotional positive response when players played the Riskio card game with a professionally printed card deck opposed to early home printed versions.  It was noticed by the players who played with the previous versions of the game, who commented quality of the professional designed and printed card decks over previous quality of the card decks. Serious cyber games can create positive and negative emotions of players. However, there is little guidance on how to create these emotional experiences~\citep{mullins2020gamification} and how gamification can enhance user engagement remains unclear~\citep{suh2018enhancing}. In the 1970s, several research pieces into intrinsic and extrinsic motivation found that students paid to perform some task were less likely to complete the task in free time \citep{sansone2000intrinsic}. A person is said to be intrinsically motivated in an activity when they receive no reward and only satisfaction in the activity and for extrinsic motivation to obtain a reward or to avoid punishment \citep{enzle1978increasing}. Cognitive evaluation theory (CET) is a sub-theory of the self-determination theory (SDT) \citep{deci2008self} and explains how individuals’ intrinsic motivations are affected by external factors~\citep{suh2018enhancing, enzle1978increasing}. In the context of gamification for cyber security awareness and education, if players feel forced by external factors, they have to play the game, affecting the players' intrinsic motivation. The work of cognitively oriented learning theorists argues the importance of intrinsic motivation~\citep{Malone:2005}. The serious game intrinsic motivation that the players value the knowledge they obtain playing the game, as can be seen, must help the players' extrinsic motivation. 
\subsection{Constructivism learning theory}
Cognitive constructivism learning theory considers that ``we construct our perspective of worlds through individual experiences and schema.'' \citep{ros2020analyzing}. That is a cognitive framework of thought or behaviour that organises categories of information and their relationships. The game players will choose their learning routes and compete against each other to reach higher levels of knowledge \citep{biro2014didactics}. There is no single constructivist theory of instruction, rather a multitude of approaches, and it can be seen as providing an alternative set of values to instructional theory~\citep{driscoll2000psychology}. Gagné's theory of instruction~\citep{driscoll2000psychology} is made up of three components: 1) Taxonomy of learning outcomes: There is a distinction between declarative and procedural knowledge. Declarative knowledge refers to the facts you remember, which is Bloom's lower order think skills~\citep{bloom1956taxonomy} as opposed to procedural, which refers to cognitive skills and Bloom's higher-order thinking skills, which should be the objective of gamification to achieve higher-order thinking. 2) Conditions for learning: The learning conditions are critically linked to the learning outcomes, and for each objective, the conditions required must be considered to achieve the objective. 3) Nine events of instruction: Gagné's nine events for instruction is the same as methods for instruction in constructivism. Not all nine instructional events need to be deployed. This would depend on the objective, but some may always be required, for example, feedback. See \autoref{tbl:Gagne9}, the nine events and Gagné belief most lessons should follow the sequence, but he recognised not absolute \citep{gagnedriscoll}.      

\begin{table}[!ht]
\centering
\begin{tabular}{|m{3cm}|m{4cm}|m{5cm}|}
\hline
\multicolumn{1}{|c|}{\textbf{Internal Process}} & \multicolumn{1}{c|}{\textbf{Instructional Event}} & \multicolumn{1}{c|}{\textbf{Action}} \\ \hline
Reception & 1. Gaining attention & Use abrupt stimulus change \tabularnewline \hline
Expectancy & \begin{tabular}[c]{@{}l@{}}2. Information learners of the\\ objective\end{tabular} &  Tell learners what they will do after learning \tabularnewline \hline
\begin{tabular}[c]{@{}l@{}}Retrieval to\\ working memory\end{tabular} & \begin{tabular}[c]{@{}l@{}}3. Stimulating recall of prior \\ learning\end{tabular} &  \raggedright Ask for recall of previously learned knowledge \tabularnewline \hline
Selective perception & 4. Presenting the content &  \raggedright Display the content with distinctive features \tabularnewline \hline
Semantic encoding & \begin{tabular}[c]{@{}l@{}}5. Providing ``Learning\\ guidance''\end{tabular} &  \raggedright Suggest a meaningful organisation \tabularnewline \hline
Responding & 6. Eliciting performance &  \raggedright Ask learner to perform \tabularnewline \hline
Reinforcement & 7. Providing feedback &  \raggedright Give information feedback \tabularnewline \hline
\begin{tabular}[c]{@{}l@{}}Retrieval and\\ reinforcement\end{tabular} & 8. Assessing performance &  \raggedright Require additional learner performance with feedback \tabularnewline \hline
\begin{tabular}[c]{@{}l@{}}Retrieval and\\ generalisation\end{tabular} & \begin{tabular}[c]{@{}l@{}}9. Enhancing retention\\ and transfer\end{tabular} &  \raggedright Provide varied practice and spaced reviews \tabularnewline \hline
\end{tabular}
\caption{Gagné's Nine Events for Instruction}
\label{tbl:Gagne9}
\end{table}
In conclusion, gamification learning theory views the learner as one of the most important actors in the learning process~\citep{biro2014didactics}. The cognitive principle of the learner motivation can be addressed by using Self Determination Theory (SDT)~\citep{deci2008self} which presents motivation as intrinsic motivation that emerges from the enjoyment of the game, and extrinsic motivation created outside of the game by workplace rules etc.~\citep{unkelos2015gamifying}. Constructivism theory could be used for instruction methods; learning objectives; and conditions for instruction to pedagogical approach to serious games design. It is using cognitive principles to create a pedagogical model for the design of serious cyber games. 

\subsection{Gaps in current design models}
This section uses three current serious games assessment models to explain the limitations in these models in creating serious cyber games and how to verify the game can achieve stated learning objectives. We select one model for further assessment using a serious cyber game called Riskio.  The first step was to select a published model to apply to Riskio to verify the serious game meets the learning objectives. Three models were considered, the Game Object Model (GOM)~\citep{amory2007game}, The Serious Game Design Assessment Framework (SGDAF) \citep{mitgutsch2012purposeful} and LM-GM Model~\citep{arnab2015mapping}. 

\begin{figure}
  \centering
  \begin{minipage}[b]{0.5\textwidth}
   \includegraphics[width=\textwidth]{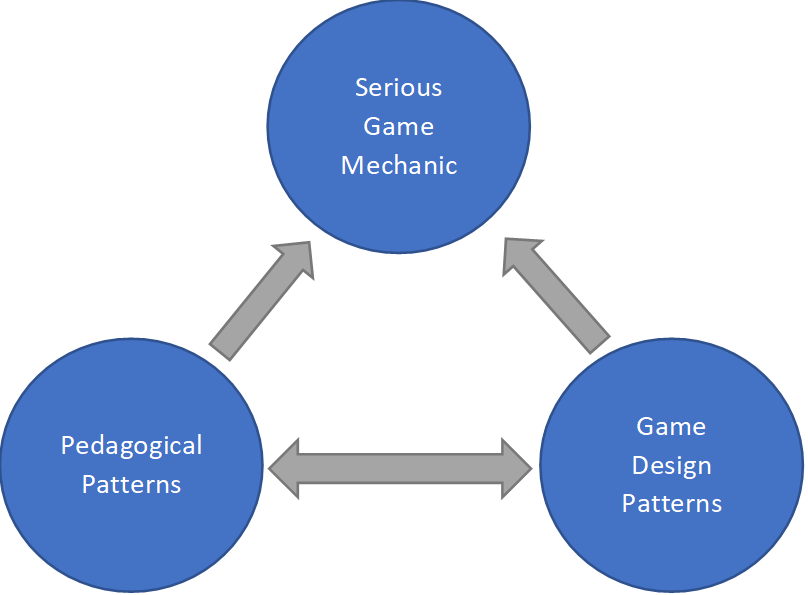}
    \caption{GOM Model}
    \label{fig:GOM_Model}
  \end{minipage}
  \hfill
  \begin{minipage}[b]{0.4\textwidth}
    \includegraphics[width=\textwidth]{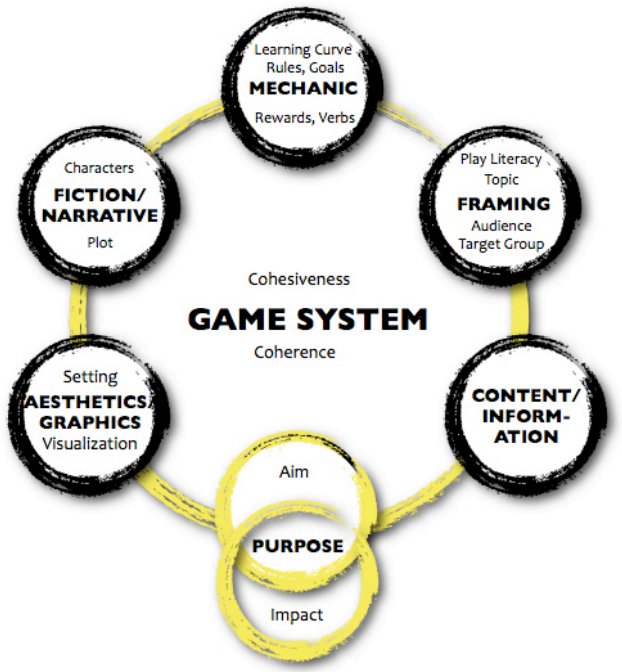}
    \caption{SGDAF Model}
    \label{fig:SGDAF_Model}
  \end{minipage}
\end{figure}

The GOM model, see \autoref{fig:GOM_Model} is based on a constructivist theoretical framework to support serious educational games development. The GOM model's five components do not show how they influence each other~\citep{arnab2015mapping} and how the components link to the game mechanics and, therefore, the serious game learning objectives. The GOM is a high-level model and will not assist in the selection of game mechanics to achieve serious game objectives and therefore, this model was excluded from the next stage testing model with Riskio game.

The SGDAF model in \autoref{fig:SGDAF_Model}, requires going through all the seven steps: 1) Purpose; 2) Content / Information; 3) Mechanic; 4) Fiction / Narrative; 5) Aesthetics Graphics; 6) Framing; and 7) Cohesiveness \& Coherence of Game. While applying the SGDAF model to Riskio, many elements were useful when reviewing against Riskio lessons learnt in the game design. For example, in step five, the aesthetics and graphics in so far, the quality of cards affected the players' enjoyment. Early versions of the game did not have professionally designed and printed cards, and players noted this in comments. This is an example where design was identified in (step 5, aesthetics \& graphics) the SGDAF Model but not in the GOM Model. However, SGDAF had a similar problem to the GOM model as in step 3 the model provides no mechanism to map the game mechanics to the learning objectives and excluded from the next stage.     

Both the GOM Model and the SGDAF Model were rejected because of the lack of methodology linking game mechanics back to the learning mechanics and the serious educational objectives of a serious cyber game. The LM-GM model was selected to create an illustrative case study using Riskio.

\begin{figure}
    \centering
    \includegraphics[width=1\linewidth]{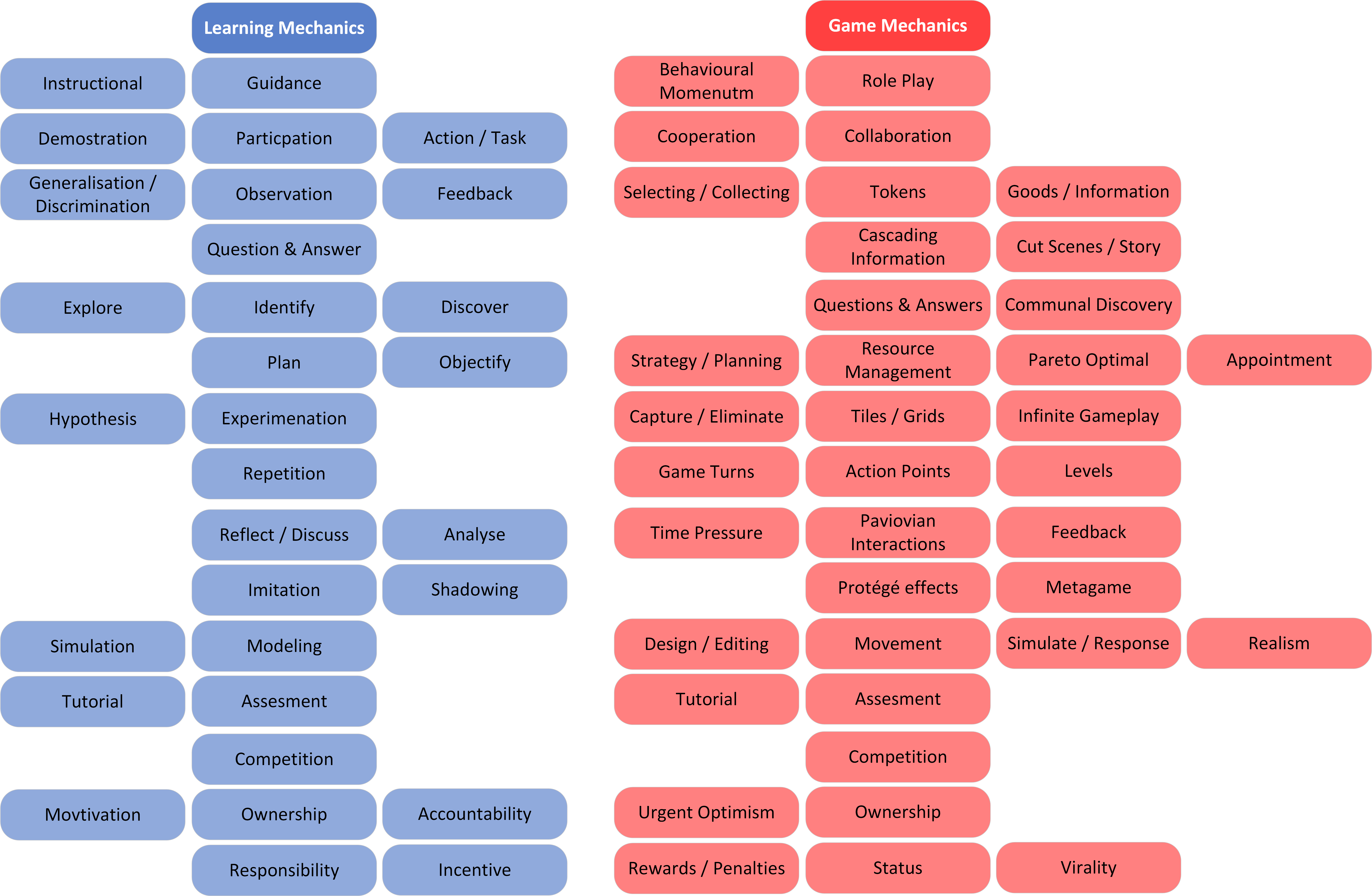}
    \caption{LM-GM Model}
    \label{fig:LM-GM_Model}
\end{figure}

The LM-GM Model, see \autoref{fig:LM-GM_Model} was created on the assumption of the fundamental design of serious games relies on the translation of learning goals into the mechanical element of the gameplay \citep{arnab2015mapping}.  The LM-GM was created to overcome the missing descriptive relationship between learning mechanics and game mechanics. The LM-GM Model also maps the game mechanics to Bloom\'s ordered thinking skills. The initial review of LM-GM Model against the components identified in the literature review showed that the LM-GM model had the potential to cover all the areas identified in the review. 

The LM-GM model is viewed as having two axes. The horizontal axis links learning mechanics analogous to game mechanics. The vertical axis run down two root nodes learning mechanics and game mechanics, each with side leaf nodes. The model is descriptive rather than prescriptive and allows users to relate learning mechanics to game mechanics. The abstract game elements (Game Mechanics) can be mapped many-to-one to the concrete game elements, and single-game Learning Objective can be achieved through different learning activities (Learning Mechanics). A single game dynamic/learning objective can be achieved through several game mechanics. Arnab et el (2014) tested the LM-GM model with a second model Amory's GOM ~\citep{amory2007game} to evaluate how effective the models were at enabling users to analyse the game. Arnab et al ~\citep{arnab2015mapping} second analysis with 26 individuals with a mixture of academics, students and game developers tested perceptions on usefulness of LM-GM and GOM models. The participants were asked to complete a suitability usability scale (SUS) which comprised of 10 questions. The overall SUS average score for LM-GM was 67.3 and GOM SUS average was 46.7. The second study with 25 final year engineering students with no gaming experience for LM-GM average SUS 71.4 and GOM average SUS 43.7. Participants found GOM easier to use than LM-GM but GOM did not identify basic pedagogical game patterns. However, the LM-GM model should be made easier to use. Participants in the Arnab et al study confirmed our view that the GOM model does not address how to implement learning objectives or map learning mechanics to game mechanics as the LM-GM does. The participants found the GOM model was much easier to understand that LM-GM model should be made easier to understand.       

The LM-GM Model was selected for the next stage to map against the Riskio game to verify the model and test with selected University staff who understand pedagogical frameworks to verify the proposed LM-GM Model using an illustrative case study.   

\begin{figure}[!ht]
	\centering
	\includegraphics[width=0.6\textwidth]{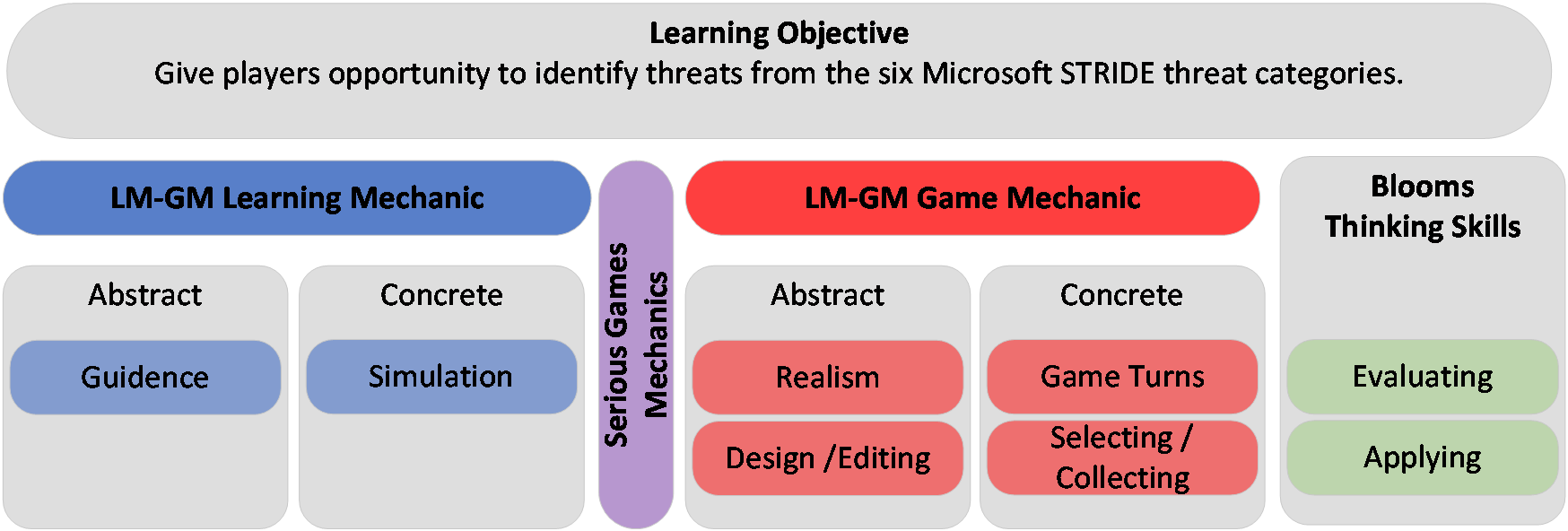}
	\caption{Links between model components}
	\label{fig:Links_LM-GM_ModelOjectiveExample}
\end{figure}

\paragraph{\textbf{Riskio LM-GM Model illustrative case study.}} We started by mapping LM-GM Model using one learning objective from the Riskio game to show the complicated relationship between learning mechanics and game mechanics. Riskio is a game where players identify threats on a game board, and we used this learning objective to map to the LM-GM model, see \autoref{fig:Links_LM-GM_ModelOjectiveExample} for an example of one Riskio game objective.

\begin{figure}[!ht]
	\centering
	\includegraphics[width=1\textwidth]{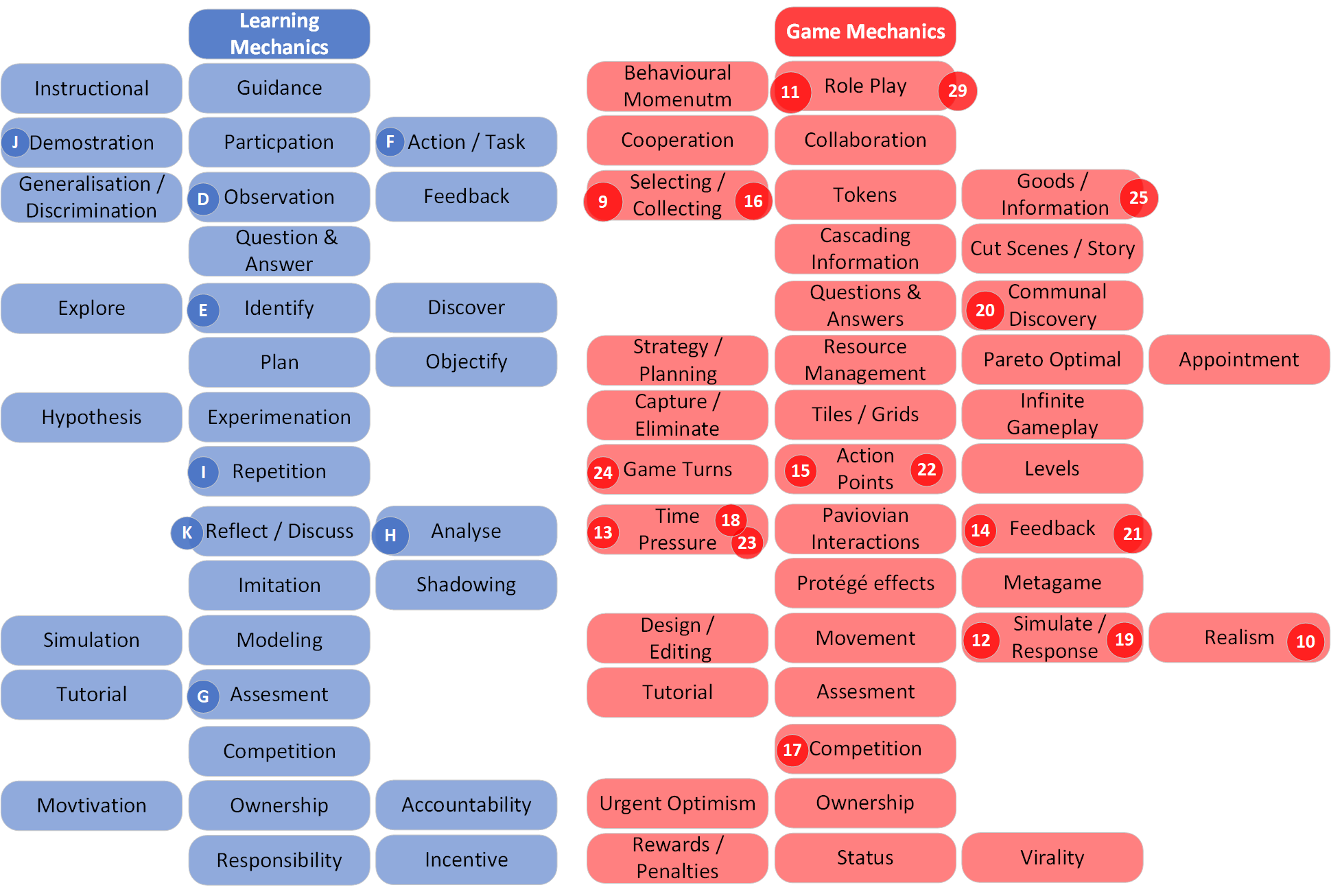}
	\caption{Riskio Game Play and Links to LM-GM Model}
	\label{fig:Links_LM-GM_MappedRiskio}
\end{figure}

\begin{figure}[!ht]
	\centering
	\includegraphics[width=0.8\textwidth]{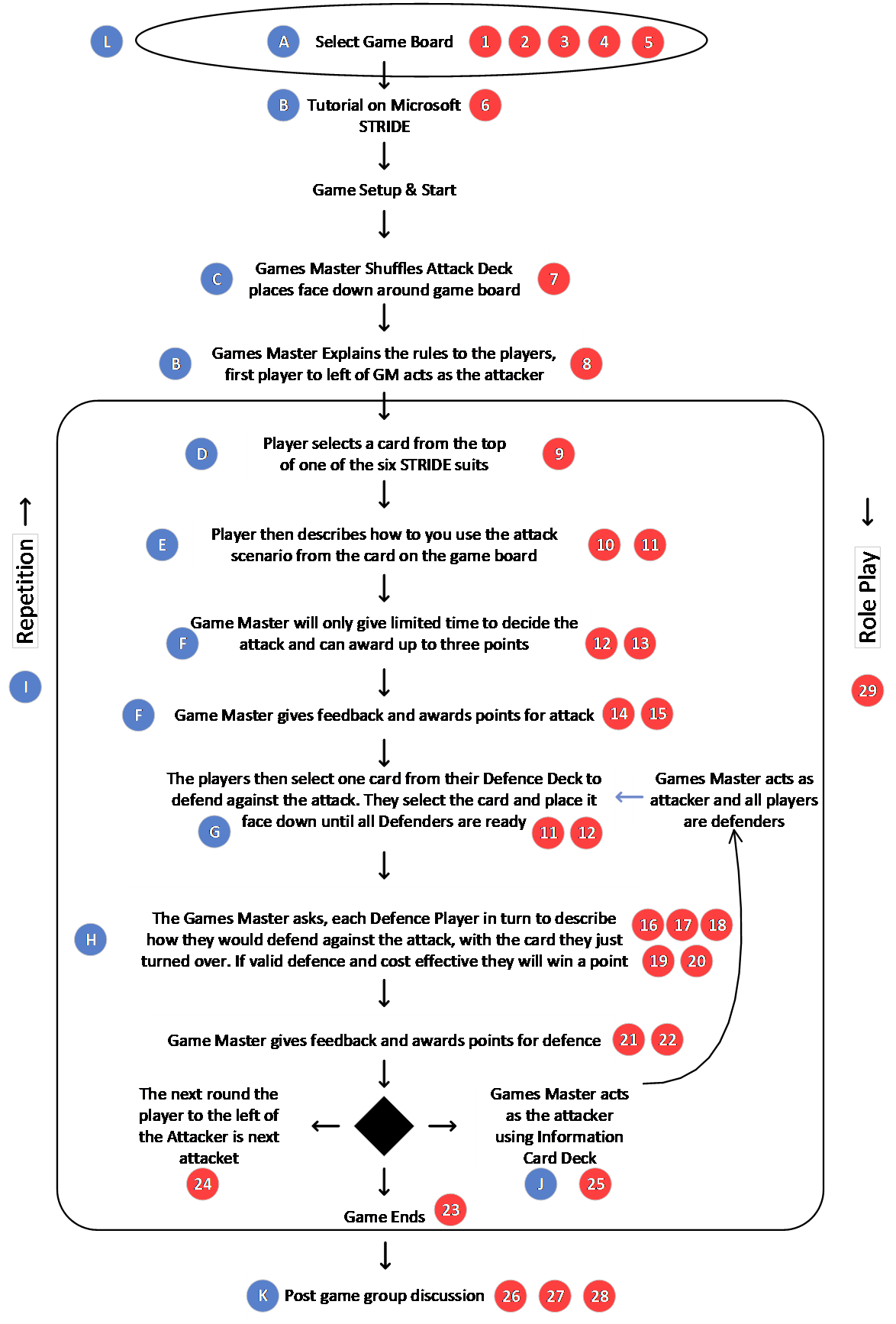}
	\caption{Riskio Game Play}
	\label{fig:Links_LM-GM_MappedRiskioGamePlay}
\end{figure}

Using the LM-GM model to test the evaluation of the Riskio game, it took several hours to refine the evaluated model before it was to an acceptable level. \autoref{fig:Links_LM-GM_MappedRiskio} which links Riskio gameplay/mechanics to the LM-GM Model and \autoref{fig:Links_LM-GM_MappedRiskioGamePlay} shows the Riskio game play/mechanics linked to the LM-GM Model. The Riskio game mapped to the LM-GM model was put into an illustrative case study to test the model's understanding with University lecturers in cyber security and cyber professionals. The feedback was they found the LM-GM model difficult to understand. There seemed to be confusion over the nodes and leafs in the model and the horizontal relationship between learning and game mechanics. Other issues noted with the mapping Riskio gameplay, see \autoref{fig:Links_LM-GM_MappedRiskio} and \ref{fig:Links_LM-GM_MappedRiskioGamePlay}, the mapping does not link the gameplay to the cognitive theory.     

\paragraph{\textbf{Conclusions on Assessment of Three Pedagogical Models.}} Some of the pedagogical models do not link game mechanics to the learning objectives, for example, GOM and SGDAF. Using the LM-GM Model, we were able to link learning objective to game mechanics and the learning objective, see \autoref{fig:Links_LM-GM_ModelOjectiveExample}. Although the LM-GM Model proved useful in mapping the pedagogical principles for serious cyber games, it took considerable time to learn the model or even explain it. The next step we try to create a model that can map pedagogical principles and link game mechanics.

\section{MOTENS Model}
\label{sec:motens_model}
\label{sec:MOTENS}
In this section, we propose a new pedagogical model called MOTENS for serious cyber games and how the model can be linked from the game mechanics to the pedagogical theory of learning.

\subsection{MOTENS Model}
\label{sec:MOTENSModel}
\begin{figure}[!ht]
	\centering
	\includegraphics[width=0.6\textwidth]{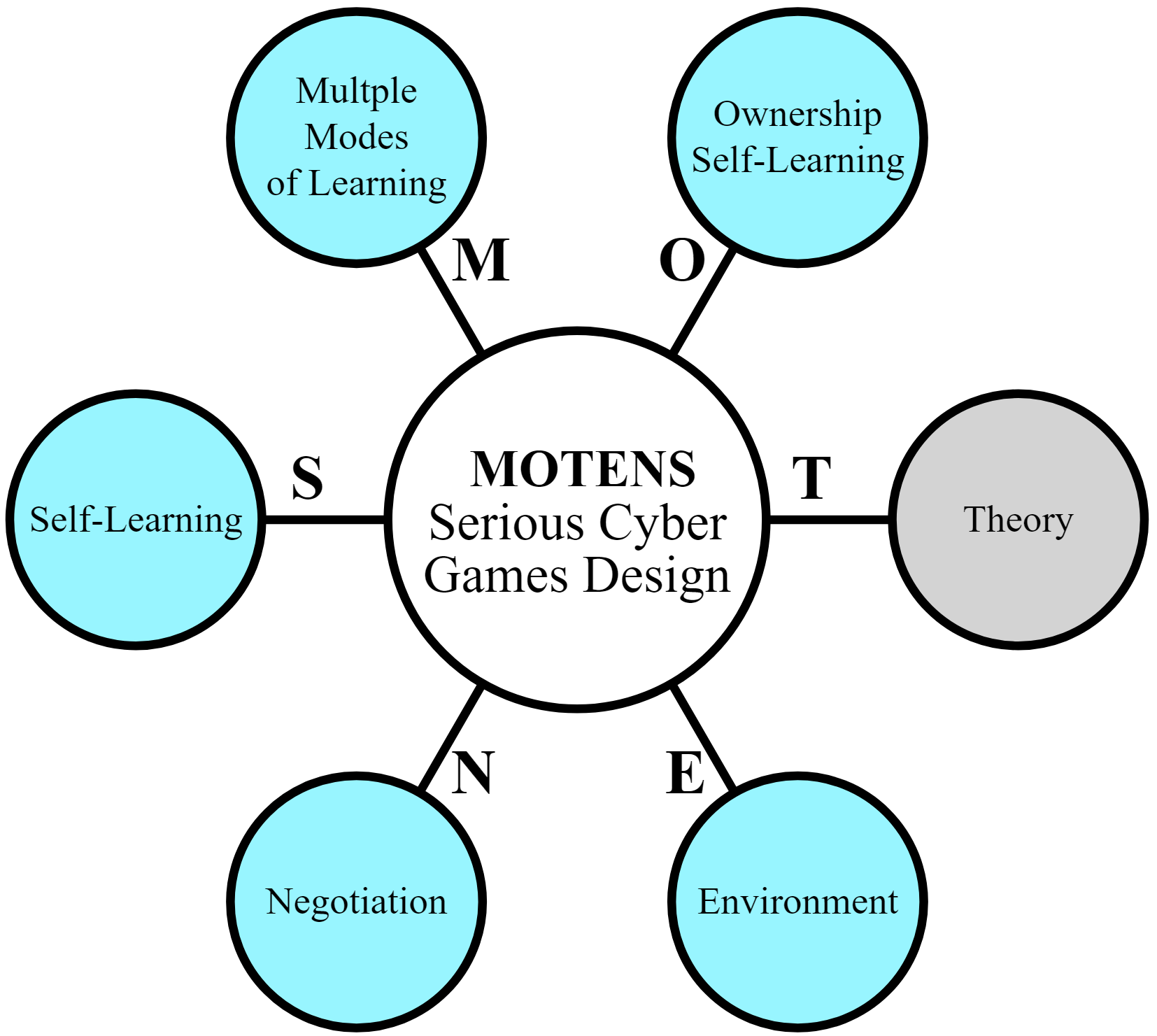}
	\caption{MOTENS Model}
	\label{fig:MOTENS_Summary}
\end{figure}

The MOTENS model, see \autoref{fig:MOTENS_Summary} was created based on the gaps identified in the current games for pedagogical for design of serious games for cyber security awareness and education and experience of lessons learnt from the creation of Riskio. The model is designed for the assessment of serious cyber games rather than other types of serious games for education.

The MOTENS model is made up of six high-level components, five of the components can be directly linked to the
games design/mechanics and one component ‘Theory’ which is the supporting the cognitive principles of the design,
including players’ motivation. 

\begin{table}[!ht]
	\centering
	\begin{tabularx}{\textwidth}{l X}
		\textbf{Multiple Modes of Learning} & The game mechanics that provide opportunities to learn a wide range of attacks and defences with players from different backgrounds.\\ 
		& \\
		\textbf{Ownership Self-Learning} & Provide different game scenarios to meet learning objectives. \\ 
		& \\	
		\textbf{Theory} & The theory that supports the design.\\
		& \\
		\textbf{Environment} & Create a game play for players in a game setting they understand and appropriate learning environment.\\
		& \\
		\textbf{Negotiation} & Change the role from teacher to coaching not lecturing and from content delivery to problem-based learning.\\	
		& \\
		\textbf{Self-Learning} & To create self-learning by use of problem-based learning; learning hierarchy; and build on players current knowledge. \\
	\end{tabularx}
\end{table}

\subsection{MOTENS Design Stages}
\label{MOTENSDesignStages}
To design and create a serious cyber game the proposed model takes you through the following design stages: 
\paragraph{\textbf{Stage 1}:} Target Game Players to identify and segment your target players into \emph{non-gamers} and \emph{gamers}. For example, Riskio primary target for game was Students, identified as gamers, with secondary group employees as non-gamers. 
\paragraph{\textbf{Stage 2}:} What game you want to create, decide category either Secure Software development or Security Awareness and Education and then decide the types of serious game: Card Games, Computer Games; Board/Table Games; or Speciality Games. Decide if your game will have a games master. 
\paragraph{\textbf{Stage 3}:} Create initial MOTENS Design/Mechanics map.  For example, it was noted that in mapping Riskio difference between Students and Employees with \emph{D11) Players Current Knowledge}. Students' intrinsic motivation to accomplish and play the game was to want a high gamification level with a random selection of the attack card category. In contrast, employees wanted to learn and low gamification and select the attack category.  
\paragraph{\textbf{Stage 4}:} Design the Game, create the game using MOTENS model. Go through five steps to design the game. Step 1, the initial design decisions. Step 2, decide the pre-game process. Step 3, design game play. Step 4, design end of game process, and Step 5 review and test gameplay. Brief examples using Riskio: Step 1, Use Microsoft STRIDE for threat model and for the defence cards NIST~\citep{NISTcyber} and NCSC~\citep{NCSC:CE} frameworks . Step 2, before playing Riskio tutorial on Microsoft STRIDE for all players. Step 3, identified different requirement to allow employees to select the attack card category. Step 4, allow time for players to discuss the game at the end of each round. Step 5, players had difficulty holding all cards, change the design to print attack category on the back of the card and only select one card at a time. 
\paragraph{\textbf{Stage 5}:} Test and Evaluate by testing the game by playing with target players and changing design/mechanics if required from player feedback. 

\subsection{Pedagogical Principles - MOTENS Model (Theory)}
In this section, we explain the pedagogical principles of the MO\textbf{\underline{T}}ENS Model. It is acknowledged that serious cyber games can provide an environment, that motivates player's to learn, however, an entertaining and fun serious game does not necessarily mean playing the game meets the objectives of the game~\citep{rooney2012theoretical}. For this reason, the pedagogical model must integrate the gameplay supported by theory to ensure the games learning objectives are met. This section explains using Riskio game how the MOTENS Model supports both the serious game design and how this is supported by pedagogical learning theory. 

\paragraph{\textbf{T1) Learning Hierarchy.}} 
Learners will sometimes memorise information without understanding the concept. For example, they may be able to recite what a Distributed Denial Service Attack (DDoS) is but not comprehend its meaning, and this would be an example of Bloom's taxonomy of lower-level thinking, \citep{bloom1956taxonomy}. Gagné's proposed a \emph{`learning hierarchy a set of component skills that must be learned before the complex skill can be learned'} \citep{gagne1992principles}. Using the DDoS example, learners understand the issues around the `availability' of systems, understand what a distributed attack is, and understand what a denial of service attack is. Learning these three components should join together to understand a DDoS attack and example in Bloom's taxonomy of higher-order thinking skills.   

\paragraph{\textbf{T2) Accountability Versus Responsibility.}} In serious games, accountability is the opposite of responsibility \citep{mayer2014research}. For players to be accountable, they must know the reason for playing the game and the effects or consequences of playing the game. In contrast, responsibility is to critically reflect on the short and long-term value and consequences for playing. \citep{mcgonigal2011reality} and this is important in getting the players thinking from Bloom's lower-order thinking skills of retention through high-order skills of understanding, applying, and evaluating to creating. For accountability the games rules and learning objectives must be clear to the players and for responsibility the players must value the learning objectives. 

\paragraph{\textbf{T3) Constructivism.}} Constructivist learning theory states that the learning process is unique by the learner and gamification theory looks at the learning process, which is from two different points of view at the same time, the first view uses individual perspective and second view from community-based learning \citep{biro2014didactics}. Any theory of instruction needs to include: methods of instruction; learning objectives; and conditions for instruction \citep{driscoll2000psychology} and consider both these perspectives.

\emph{Methods of Instruction.} The constructivism theory is based on the belief that learning occurs as learners are actively involved in the process of meaning and knowledge construction as opposed to passively receiving information \citep{fosnot:1996}, constructivist-oriented approach concentrates on the learners constructing their understanding during social interactions \citep{maor1999teacher}. Therefore, the use of gamification, to teach cyber security awareness and education the game must promote the interactions that increase the discourse and personal construction.  In the design of Riskio game we followed the principles of constructivism \citep{hart2020riskio}, which has been the predominant learning theory used in education programs for young children, college and university students \citep{fosnot:1996, rolloff2010constructivist}. 

In a constructivist learning environment, learners work primarily in groups and learning and knowledge are interactive, and facts and knowledge change with experience \citep{bada2015constructivism}. There are a significant focus and emphasis on social and communication skills and collaboration and exchange of ideas. This is contrary to the traditional learning environments where learners work primarily alone. Learning is achieved through repetition. The subjects are strictly adhered to and guided by a textbook. 
Driscoll \citep{driscoll2000psychology} summarised the five conditions for instruction for constructivism are: (C1) complex and relevant learning environment; (C2) social negotiation; (C3) multiple perspectives and multiple modes of learning; (C4) ownership in learning; and (C5) self-awareness and knowledge construction, see \autoref{tbl:ConstructivistRiskioExample} an example links between, constructivist five conditions, MOTENS Model and Riskio game mechanics.   

\begin{table}[!ht]
	\centering
		\begin{tabular}{|p{4cm}|p{4cm}|p{4cm}|}
			\hline
			\multicolumn{1}{|c|}{\textbf{\begin{tabular}[c]{@{}c@{}}Constructivist\\ Conditions\end{tabular}}} & \multicolumn{1}{c|}{\textbf{Link to MOTENS Model}} & \multicolumn{1}{c|}{\textbf{Example in Riskio}} \tabularnewline \hline
			\raggedright C1: Complex and relevant learning & \raggedright Environment: D5) Security Defences & \raggedright Players using Riskio defence deck can learn complex and basic defences \tabularnewline \hline
			\raggedright C2: Social negotiation & \raggedright Negotiation: D10) Opportunity to Discuss Game Play & \raggedright End of each round of attack and defence games master encourages players to discuss and learn from other players \tabularnewline \hline
			\raggedright C3: Multiple perspectives and multiple modes of learning & \raggedright Multiple Modes of Learning: D1) Game Mechanics & \raggedright Each Riskio game board can provide multiple metaphors and analogies and multiple interpretations, the hallmark of Cognitive Flexibility Theory \citep{spiro2003cognitive} this can be done in Riskio with different case studies that support the game board to give different perspectives  \tabularnewline \hline
			\raggedright C4: Ownership in learning & \raggedright Ownership Self-Learning: D2) Different Game Scenarios & \raggedright Riskio can be played with different game scenarios, having contextualised game objectives players are encouraged to self-learn, however concern not all students achieve "buy-in" \citep{perkins1991constructivism}  \tabularnewline \hline
			\raggedright C5: Self-awareness and knowledge construction & \raggedright Negotiation: D7) Role of Games Master & \raggedright Games master helping players become aware of thinking process, what theorist call metacognition \citep{driscoll2000psychology} \tabularnewline \hline
			\end{tabular}%
	\caption{Constructivist Conditions linked to MOTENS with Riskio Example}
	\label{tbl:ConstructivistRiskioExample}
\end{table}

\emph{Game Learning Objectives.} The constructivist approach to identifying the learning goals, emphasises the learning context it is not to assure that students know particular things, but rather to show them how to construct plausible interpretations \citep{bednar1992theory}. Using the constructivist approach the following objectives were identified for the pedagogical model for serious cyber games: \emph{Transfer Knowledge (TK)} - Applying knowledge to other acquired skills; \emph{Serious Games Types (SGT)} - The model can be used for desired game types; \emph{Authentic Learning (AL)} - Linked to real-world learning; \emph{Ownership Self-Learning (OSL)} - Active Self learning; \emph{Game Environment (GE)} - Game scenarios appropriate learning environment; \emph{Intrinsic Motivation (IM)} - Ability to play game (Gamers) or provides learning (non-gamers);  and \emph{Extrinsic Motivation (EM)} - Knowledge that is valued.

\emph{Conditions of Instruction.} We proposed using hypermedia designs, collaborative learning, problem scaffolding, and problem-based learning to create constructivist conditions for instruction. Examples of how implemented in Riskio game: \emph{Hypermedia Designs} the use of game boards that are a small but complete subset of real environments; \emph{Collaborative Learning} provide an opportunity for players to discuss at the end of each round the attack and various different defences used; \emph{Problem Scaffolding} interactions between the payers and the games master can provide different levels of support and \emph{Problem Based Learning} case studies supporting games board can state problems that players need to decide how to defend.     

\emph{Link to Gagné Nine Instructional Events.} The nine instructional events can be mapped to the MOTENS model, using Riskio gameplay as an example. 1) gaining attention, 2) Inform learners of objective: Tutorial at start of the game; 3) Simulate recall of prior learning: Discussion at the end of each round and end of the game; 4) Presenting content: Using graphics and icons on game board understood by players; 5) Proving learning guidance: Using games board that are relevant to the players; 6) Eliciting performance: Players try to find most economical defence; 7) Proving feedback: Games master giving feedback on attacks and defence; 8) Assessing performance: and 9) Enhancing retention and transfer: Games master can act as an attacker using information deck.      
\paragraph{\textbf{T4) Gamification.}} Gamification is about applying game mechanics to non-gaming activities, for example, training to make the activity more engaging \citep{routledge2016games}. Serious games use these techniques to provide a fun, enjoyable educational environment where the game participants learn by playing the game. Gamification does not mean game design requires designers to concentrate on competitive features in the design between players. Studies have proven a positive influence of serious games using gamified cooperation to create meaningful connections amongst players, and it facilitates similar learning, and motivational outcomes as gamified competition \citep{dindar2021experimental}. In MOTENS design stage 3 (see \autoref{MOTENSDesignStages}) creates a design/mechanics map to consider selecting the correct gamified content for the game to meet target players requirements.  

\begin{figure}[!ht]
	\centering
	\includegraphics[width=0.8\textwidth]{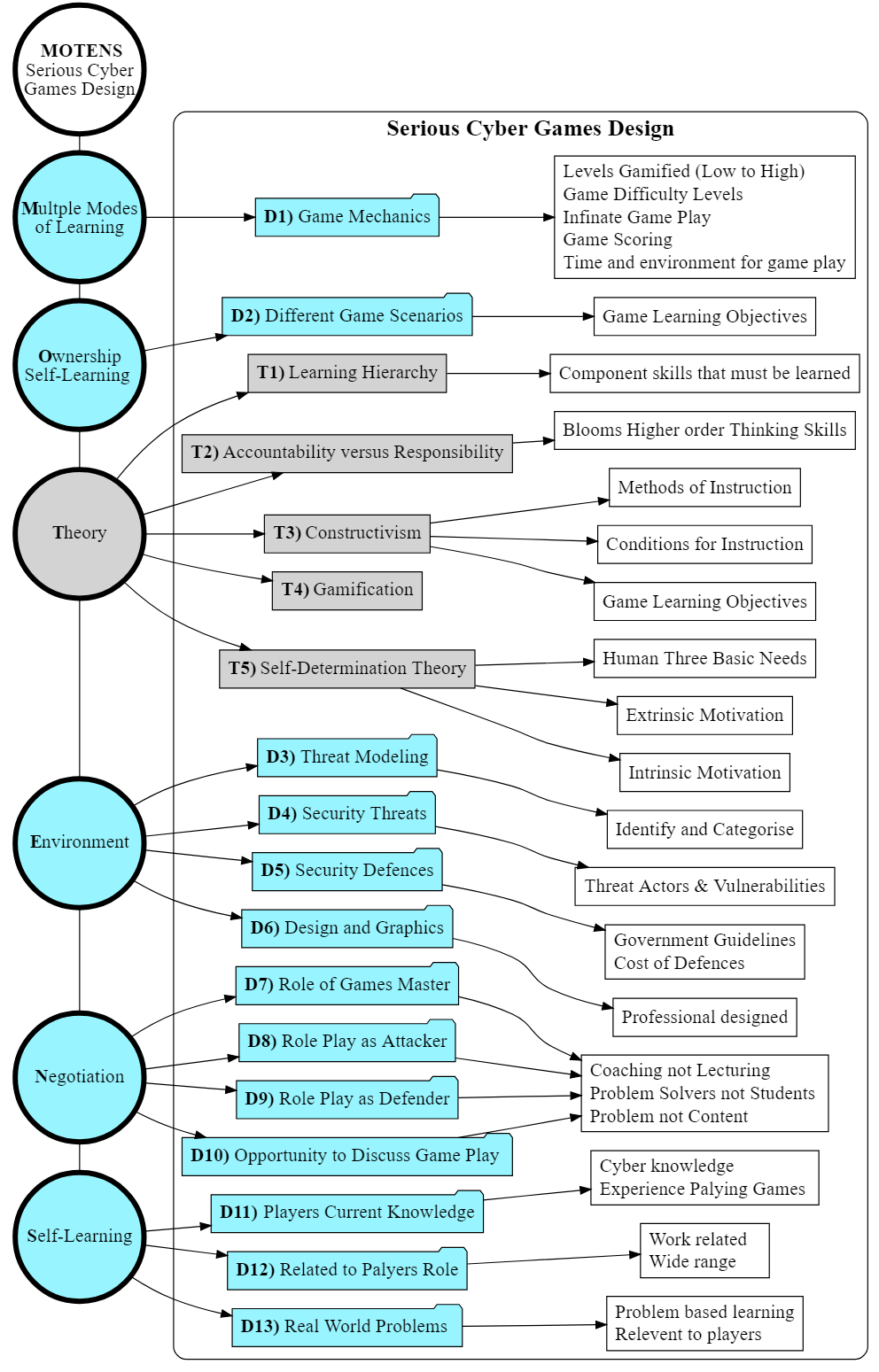}
	\caption{MOTENS Model Linked to Theory and Game Mechanics}
	\label{fig:MOTENS_detail}
\end{figure}

\paragraph{\textbf{T5) Self-Determination Theory.}} Deci and Ryan \citep{deci2008self} proposed a macro-theory called Self-Determination Theory (SDT). although the work started on SDT in the 1970s it can be applied today to gamification. SDT presents motivation as extrinsic motivation, the external factors, and intrinsic motivation, the internal factors. SDT also presents three basic psychological needs: Competence - Can perform activity well; Autonomy - Feeling you are in control; and Relatedness - Sense of belonging. To be intrinsically motivated all three needs to be satisfied and extrinsic motivation needs at least competence and relatedness must be satisfied \citep{conejo2019detailing}. Conejo \& Hounsell \citep{conejo2019detailing} propose a modification to the existing framework to assist designers of games. They suggest that some game design frameworks address motivation superficially, while others focus exclusively on motivation. \autoref{tbl:SDTRiskioExample} shows example of links between SDT theory, MOTENS Model, see \autoref{fig:MOTENS_detail} and the Riskio game.

\begin{table}[!h]
	\centering
		\begin{tabular}{|p{3cm}|p{4.5cm}|p{4.5cm}|}
			\hline
			\multicolumn{1}{|c|}{\textbf{\begin{tabular}[c]{@{}c@{}}SDT Theory\end{tabular}}} & \multicolumn{1}{c|}{\textbf{Link to MOTENS Model}} & \multicolumn{1}{c|}{\textbf{Example in Riskio}} \tabularnewline \hline
			\raggedright Competence & \raggedright Multiple Modes or Learning: D11) Players Current Knowledge & \raggedright Riskio game difficulty levels, enables all players to be able to find attacks and defences \tabularnewline \hline
			\raggedright Autonomy & \raggedright Negotiation: D8) Role Play as Attacker; D9) Role Play as Defender & \raggedright Players are able to select the category of attack and defence cards \tabularnewline \hline
			\raggedright Relatedness & \raggedright Ownership Self-Learning: D2) Different Game Scenarios & \raggedright Games boards can be changed to relate to players work learning objectives\tabularnewline \hline
			\end{tabular}%
	\caption{SDT linked to MOTENS with Riskio Example}
	\label{tbl:SDTRiskioExample}
\end{table}

\subsection{MOTENS Game Mechanics}
In this section, we explain the five components of the \textbf{\underline{MO}}T\textbf{\underline{ENS}} Model linked to games design/mechanics. 

\paragraph{\textbf{D1) Game Mechanics.}} In design stage 1 of the design you segment the players into gamers and non-gamers and important that you select the correct level gamified content for target players. For example, students (identified as gamers) might want random selection of threat category by throwing a dice, whereas non-gamers will want to select threat category. 

\paragraph{\textbf{D2) Different Game Scenarios.}} You can select different game scenarios, for example in Riskio \citep{hart2020riskio} we used University Fees Office as this proved most popular with player but could use network diagrams or other fictional settings.

\paragraph{\textbf{D3) Threat Modeling.}} This is where you select a threat model for example in Riskio we used Microsoft STRIDE as suited our learning objectives, but you might want to use different model for example serious game about hardware supply chain use CIST \citep{halak2021hardware} a threat model created for hardware supply chain.

\paragraph{\textbf{D4) Security Threats.}} The game must expose the players to the most common threats and for example for Riskio we identified in cyber security reports (e.g. by SANS and Symantec), security guidance (e.g. by NCSC or NIST), and security practices (e.g. by OWASP).

\paragraph{\textbf{D5) Security Defences.}} The game defences should be based on a wide range of attacks and countermeasures. Published frameworks should be used to build on players knowledge for example: NCSC Cyber Essentials \citep{NCSC:CE}, and 10 Steps to Cyber Security \citep{NCSC:10}.

\paragraph{\textbf{D6) Design and Graphics.}} In design stage 4 you should not only test the game mechanics and game play but verify the design and graphics. The quality of design of logos, cards and icons etc. can affect the players’ enjoyment of the game and this was noted in playing early version of Riskio with home printed cards. All graphics should be professionally designed, created, and high quality printed where required. 

\paragraph{\textbf{D7) Role of Games Master.}} If the game has a games master, their role is not to provide knowledge to the learners, but to prompt and facilitate discussion. This can be done by designing stages in game play that facilitate discussion. The games master encourage learners inquiry by asking thoughtful, open-ended questions and encouraging learners to ask questions of each other; seek elaboration of learners’ initial responses; encourage learners to engage in dialogue, both with the games master and with one another (C4: Ownership in learning, see \autoref{tbl:ConstructivistRiskioExample}).

\paragraph{\textbf{D8) Role play as Attacker.}} The design should convey the breadth of vulnerabilities and attack methodologies that can be exploited by attackers.

\paragraph{\textbf{D9) Role play as Defender.}} The design should improve the understanding of the diversity of possible countermeasures that can be considered to prevent, detect, or mitigate cyber attacks. Players should learn the different defence strategies and not all attacks can be prevented.  

\paragraph{\textbf{D10) Opportunity to discuss game play.}} The game should enable the players to cooperate and prompt and facilitate discussion about attacks and defences to create meaningful connections amongst players \citep{dindar2021experimental}.

\paragraph{\textbf{D11) Players current knowledge.}} The selection of some of game mechanics can build on players current knowledge. For example, if they already use a threat model consideration to use this in games design to build on players knowledge.

\paragraph{\textbf{D12) Related to Players Role.}} The designer can choose between creating a serious game where player plays a work-related role or can be given a specific role, for example as attacker.  

\paragraph{\textbf{D13) Real World Problems.}} The emphasis on solving should be on real-world problems \citep{seng2004students} and move problem based learning \citep{seng2000reflecting}. 

\subsection{Assessment of MOTENS Model.} 
\begin{figure}[!ht]
	\centering
	\includegraphics[width=0.9\textwidth]{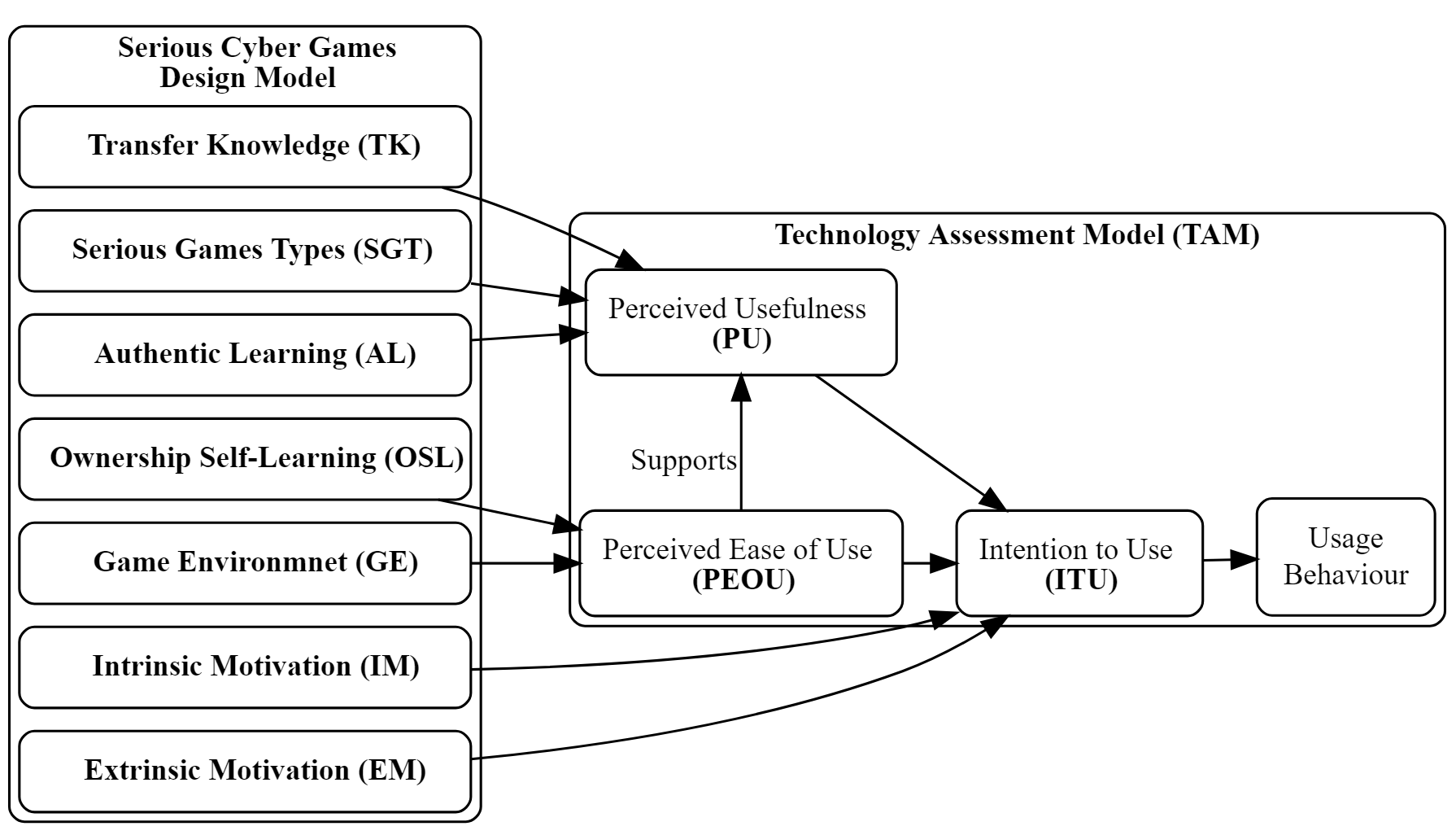}
	\caption{TAM Model linked to Assessment of MOTENS Model}
	\label{fig:TAMModel}
\end{figure}

The next stage is to develop the MOTENS model into an illustrative case study and then test the case study with cyber professionals in higher education, researchers in cyber security and cyber professionals to test the model. The testing of perception will be for perceived ease of use (PEOU); perceived usefulness (PU); intention to use (ITU); and efficacy of the new model, see \autoref{fig:TAMModel} which is based on TAM model by Yusoff~\citep{yusoff2010conceptual}. The TAM Model in \autoref{fig:TAMModel}, maps serious games learning objectives to the TAM Model. Studies have revealed a lack of methods of measuring the effectiveness of information security training \citep{nguyen2020design}, and many surveys ask questions focusing mainly on the knowledge and not the change in behaviour \citep{khan2011effectiveness}. The case study will be able to verify the MOTENS Model has the potential to assist in the design of serious cyber games. However, there is still a lack of psychological theories to demonstrate increasing players' knowledge and changing security behaviours which is the ultimate objective of any security awareness and education training program. Using the MOTENS Model to design serious cyber games can link player motivations using SDT and use constructivism to link instruction; learning objectives; and conditions for instruction. The last part in testing the game with players can link serious game objectives to the TAM Model to verify the perceived ease of use, perceived usefulness, and intention to use, see \autoref{fig:TAMModel}.     

\section{Illustrative Case Study for Efficacy of MOTENS}
\label{sec:case_study}
This section sets out the illustrative case study used to evaluate the MOTENS model. The case study's goal was to demonstrate that the proposed framework is effective to design serious cyber games. The target audience for the case study is not users who would play serious cyber games rather people working in universities, research, PhD students and cyber educators who would be involved in designing serious cyber games for awareness and education or working in research of cyber security.  

\subsection{Study Design}
The design of the study based on the Technology Acceptance Model (TAM) that uses three constructs predict the user acceptance of new technology \citep{Davis:1989, yusoff2010validation}. 1) \emph{perceived ease of use} (PEOU), the degree to which a person believes that using a particular technology is free of effort; 2) \emph{perceived usefulness} (PU), person’s subjective probability that using a particular system would enhance his or her job performance; 3) \emph{intention to use} (ITU), the extent to which a person intends to use a particular system. The participants were also asked three background questions and two questions on the level of expertise in cyber security technologies and cyber security  awareness and education.  

The participants were emailed a link to a short video explaining the MOTENS model and given a case study explaining the background of the MOTENS model and using Riskio serious cyber game \citep{hart2020riskio} as an example how the model can be applied to create a serious cyber game. The participants' were then asked to complete a questionnaire with five background questions, see \autoref{tbl:QuestionsBackground} and eight on their perceptions of the MOTENS serious games design model, see \autoref{tbl:QuestionnaireMOTENS}.  

\begin{table}[!ht]
	\centering
		\begin{tabular}{|c|p{10cm}|}
			\hline
			 \multicolumn{2}{|c|}{\textbf{Background}} \\ \hline
			\raggedright Q1 & \raggedright Which team/function area do you work in at your organisation? \tabularnewline \hline
			\raggedright Q2 & \raggedright What is your knowledge of Riskio game? (Tick all that apply) \tabularnewline \hline
			\raggedright Q3 & What is your interest in serious cyber security games? (Tick all that apply) \tabularnewline \hline
			 \multicolumn{2}{|c|}{\textbf{Expertise}} \\ \hline
			\raggedright Q4 &  \raggedright How would you describe your level of expertise in cyber security technologies? \tabularnewline \hline
			\raggedright Q5 & \raggedright How would you describe your level of expertise in cyber security awareness and education? \tabularnewline \hline
			\end{tabular}%
	\caption{Participant Background \& Expertise Questionnaire}
	\label{tbl:QuestionsBackground}
\end{table}

\subsection{Case Study Questionnaire} 
The questionnaire see \autoref{tbl:QuestionnaireMOTENS} was used to collect impressions about the proposed MOTENS pedagogical design model for serious cyber games. For the analysis the results were aligned to 1 being negative answer and 5 being positive answer. The hypothesis was formulated using the TAM model as follows to evaluate the MOTENS model:
\begin{itemize}
    \item \emph{PU}: The participants found the MOTENS model covered all types of serious cyber games and useful (Q1, Q2). 
    \item \emph{PEOU}: The participants found that the MOTENS model would be easy to use and able to use it to match to learning objectives (Q3, Q4).
    \item \emph{PU}: The participants agree the value in using MOTENS to design serious cyber games to achieve desired learning outcomes (Q5).
    \item \emph{PU}: The participants agree the value in using serious cyber games to achieve desired learning outcomes (Q6).
    \item \emph{ITU}: The participants would use or recommend the use of MOTENS to design serious cyber games (Q7, Q8). 
\end{itemize} 

\begin{table}[!ht]
	\centering
		\begin{tabular}{|c|c|p{10cm}|}
			\hline
			\textbf{Q} & \textbf{Category} & \multicolumn{1}{c|}{\textbf{Question}} \tabularnewline \hline
			\multicolumn{3}{|c|}{\textbf{Serious Cyber Games Types (SGT)}} \\ \hline
			\raggedright Q1 & PU & \raggedright I found the model covered all types of serious cyber games I expected. \tabularnewline \hline
			\raggedright Q2 & PU & \raggedright Using the I feel that the model would be useful in the design of the all the types of games: Card Games; Computer Games; Board Games; and Speciality Games (Education \& awareness only for this type), see Figure I5 in Case Study. \tabularnewline \hline
			\multicolumn{3}{|c|}{\textbf{Games Environment (GE)}} \\ \hline
			\raggedright Q3 & PEOU & \raggedright I found the MOTENS model would be easy to use. \tabularnewline \hline
			\raggedright Q4 & PEOU & \raggedright I found the MOTENS model was able to match learning objectives to serious games mechanics.\tabularnewline \hline
			\multicolumn{3}{|c|}{\textbf{Authentic Learning (AL)}} \\ \hline
			\raggedright Q5 & PU & \raggedright I think using the MOTENS model to design serious cyber games will improve learning outcomes and give greater chance to meet desired learning objectives of serious cyber games design. \tabularnewline \hline
			\multicolumn{3}{|c|}{\textbf{Transfer Knowledge (TK)}} \\ \hline
			\raggedright Q6 & PU & \raggedright I feel playing serious cyber games is an effective method to teach cyber security awareness and education and secure software development. \tabularnewline \hline
			\multicolumn{3}{|c|}{\textbf{Intrinsic Motivation  (IM) \& Extrinsic Motivation (EM)}} \\ \hline
			\raggedright Q7 & ITU & \raggedright I would recommend the MOTENS model to anyone designing a serious cyber game. \tabularnewline \hline
			\raggedright Q8 & ITU & \raggedright Overall, I think the MOTENS model will be useful to design cyber games to meet intended objectives and I would use it to help to design serious cyber games. \tabularnewline \hline
			\end{tabular}%
	\caption{Questionnaire MOTENS Serious Cyber Games Design Model}
	\label{tbl:QuestionnaireMOTENS}
\end{table}

\subsection{Threats to Validity} 
In this section, we discuss the main threats to the validity of our case study: construct, reliability, internal and external validity \citep{Wohlin:2012}.

\paragraph{Construct validity} This aspect is to what extent the research questions represent what was in mind. The threat identified was anyone in the target participants who don't feel playing serious cyber games is an effective method and will have a negative bias in the MOTENS questions. This was mitigated by asking one generic question on the effectiveness of serious games to teach cyber security awareness and education (see question 6, \autoref{tbl:QuestionnaireMOTENS}).      

\paragraph{Reliability} This is the aspect concerned with the extent to which the data and the analysis are dependent on the specific researchers. The participants were required to have an overview of the MOTENS model before they answered the questionnaire. The identified risk that the MOTENS presentation may vary in content and delivery even from the same presenter and affect the participants' questionnaire's answers. This was mitigated by recording the MOTENS presentation to ensure it was independent of the researcher presentation.      

\paragraph{Internal validity} This is of concern when causal relations are examined whether one factor investigated is a risk that the investigated factor is also affected by a third factor. We identified that participants who played the Riskio game used as an example in the MOTENS case study might be biased and to mitigate against this we asked a question about participants knowledge of Riskio game, and so any possible bias could be analysed in the results.

\paragraph{External validity} This is concerned with the extent to which it is possible to generalise the findings beyond the case study settings. A potential threat could have been to select the wrong people to participate in the study. Although the questionnaire was anonymous, this was mitigated by asking a background question and excluding any questionnaires that don't meet the target audience.

\subsection{Analysis of Case Study}
\begin{figure}[!ht]
\centering
\includegraphics[width=0.95\textwidth]{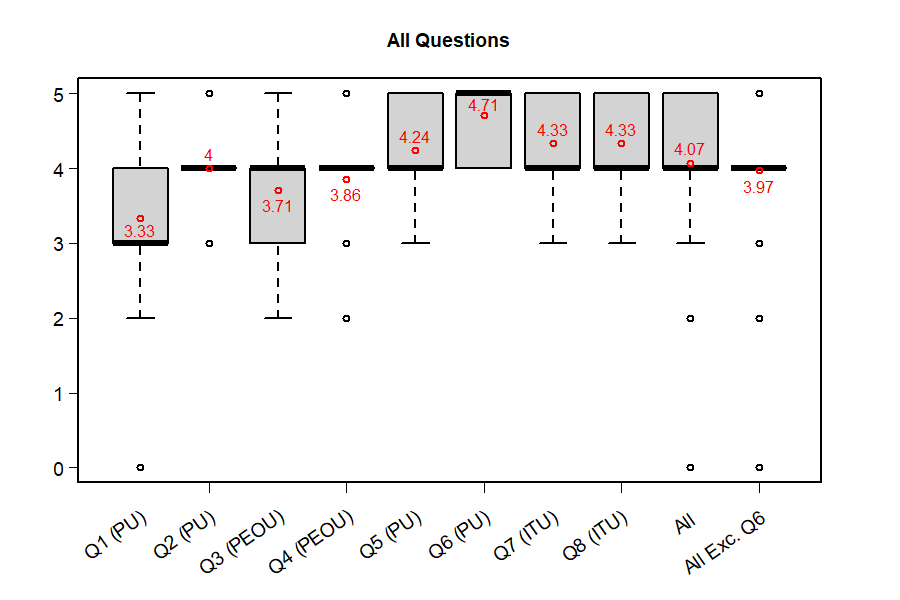}
\caption{All Questions post review MOTENS Model}
\label{fig:AllQuestions1to8}
\end{figure}

For the analysis the results of the questions from \autoref{tbl:QuestionnaireMOTENS} responses have been aligned from 1 the lowest participant perception (strongly disagree) to 5  the highest participant perception (strongly agree) and displayed in \autoref{fig:AllQuestions1to8} in a box-plot diagram and the outliers plotted as individual points. The calculation for the outliers was based on Interquartile Range (IQR) to set the minimum and maximum values to be considered: 
\[IQR = Q_{3} - Q_{1}\] 
\[Q_{1}-1.5 IQR \qquad Q_{3}+1.5 IQR\] 
\paragraph{Background Participants.} The case study involved 21 respondents, 7 working as university professor, associate professor or lecturer, 2 working in University Research Department, 5 PhD students, 1 MSc student and 6 from commercial organisations working in cyber awareness and education. Only 9 of the participants had read the Riskio published paper \citep{hart2020riskio}. The participants were asked two background questions to scale their expertise in cyber technologies and awareness and education. Both questions had similar means, expertise technologies 3.6 and expertise in education 3.2.    

\paragraph{Threats to Validity} Question 6 used to test the construct validity that participants have a general perception of the use of serious games with a mean score of 4.71. All respondents were scoring either 4 or 5 as a clear indication all participants have a high perception of benefits using serious cyber games. Background question on knowledge of Riskio game was used to test internal validity and the mean score of all eight questions in \autoref{fig:AllQuestions1to8} of the 9 participants with knowledge of Riskio game mean score was 4.1 and the 12 respondents with no knowledge mean score of 4.0 and showed no significant difference.  All the respondents met the required target for participation in the case study and no identified risks to external validity. 

\begin{figure}[!ht]
\centering
\includegraphics[width=0.9\textwidth]{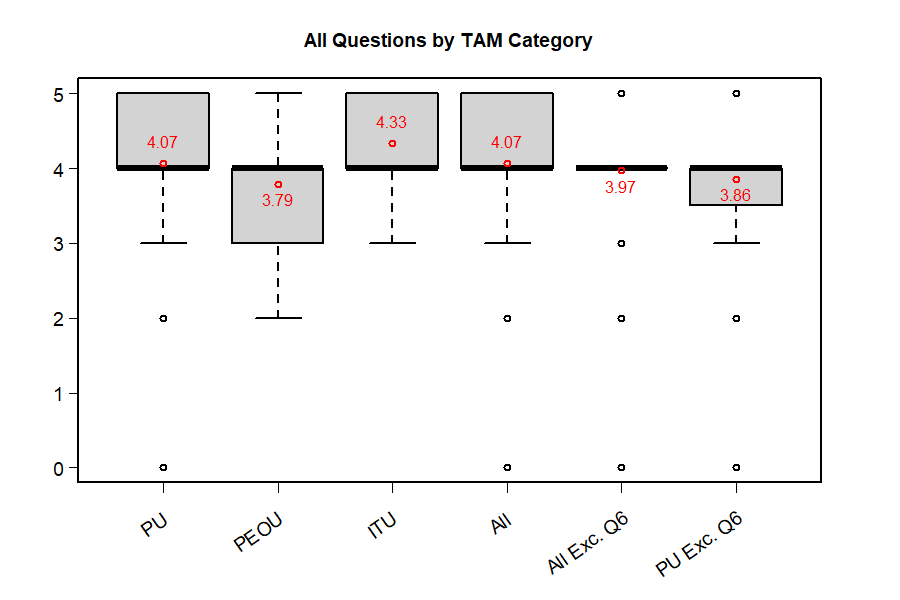}
\caption{MOTENS Questions by TAM Category}
\label{fig:TAMCategory}
\end{figure}

\paragraph{Overall Perception.} The results show the overall perception mean 4.07. There were some outliers whereas example one participant on question 1 on the MOTENS model covering all types of games commented ``I did not come into the exercise with an expectation''. The overall interquartile range (IQR) for all 8 questions being between 4 and 5. Excluding question 6, which was the generic question to test validity, the mean score was 4.71 and same IQR between 4 and 5.  

\paragraph{Perceived Ease of Use.} The mean score for questions 3 and 4 to test the PEOU was the lowest at 3.71. Question 3 regarding MOTENS model would be easy to use had the broadest range of answers with an IQR between 3 and 4. Question 4 on MOTENS's ability to match learning objectives to game mechanics the participants having similar positive perception with a mean score of 3.86 with the IQR of 4. We can conclude that participants can see how MOTENS can match learning objectives to serious games mechanics but feel it may not be easy to use.        

\paragraph{Perceived Usefulness.} The perceived usefulness of MOTENS model mean score was 4.07, including generic question 6. The mean was 3.97 when excluding generic question 6, the IQR range was between 4 and 5. The participants' feedback was they thought the MOTENS model could be useful in the creation of serious cyber games. 

\paragraph{Intention to Use.} The overall intention to use mean score was 4.33 with all participants scoring either 4 or 5 for both the question 7 on a recommendation to use and question 8  useful to meet intended objectives. Feedback from one external University proposed they would consider using the model for MSc students developing cyber games.     

\subsection{Conclusion}
The IQR for PU and ITU was 4 to 5, whereas IQR PEOU was 3 to 4 (see \autoref{fig:TAMCategory}), and we decided to create a second case study and target students who would use the model to design serious cyber games to test the PEOU in a comparative case study against another model to test the PEOU against another model. 
\section{Comparative Case Study}
\label{sec:case_study2}
This section uses a quasi-experimental comparison between the LM-GM and our new proposed MOTENS model. This study was to verify the MOTENS model ease of use by establishing the game mechanics' pedagogical intent in the model. This study was targeted at students who are currently designing serious cyber games or interested in designing serious cyber games. 

\subsection{Study Design \& Questionnaire}
To evaluate the LM-GM against the MOTENS model, we used both models to map the Riskio gameplay. However, the LM-GM model has no concept of numbering notation. In the evaluation of Arnab et al. \citep{arnab2015mapping}, the participants used their numbering to map from the gameplay they identified back to the relevant model they were evaluating. The MOTENS model components are numbered, and to ensure validity, we added a notation to the LM-GM model for the comparison to MOTENS. The experiment involved an online presentation to MSc students on serious game design models of the LM-GM and MOTENS model using the Riskio game to apply the model principles and other presentations. The participants were then given an extract of Riskio gameplay applied to both models and given a questionnaire to score each model. As with the illustrative study in \autoref{sec:case_study} we used the Technology Acceptance Model (TAM) to test the difference between the new MOTENS and the LM-GM model. Participants were asked three questions about each model (see \autoref{tbl:QuestionnaireLMGMMOTENS}). Q1: \emph{perceived usefulness} (PU); Q2: \emph{perceived ease of use} (PEOU); and Q3: \emph{intention to use} (ITU). The participants scores were aligned 1 to 5, and this enabled us to compare with the illustrative case study for MOTENS in \autoref{sec:case_study}. 

\begin{table}[!ht]
	\centering
		\begin{tabular}{|c|c|p{10cm}|}
			\hline
			\textbf{Q} & \textbf{Category} & \multicolumn{1}{c|}{\textbf{Question Asked for both LM-GM and MOTENS}} \tabularnewline \hline
			\raggedright Q1 & PU & \raggedright It will be useful to use the model to help designing serious cyber games that are effective for players to learn desired cyber educational objectives. \tabularnewline \hline
			\raggedright Q2 & PEOU & \raggedright Using the model to design serious cyber games, I can see it will be easy to map gameplay to the game mechanics and support the pedagogical educational theory and learning, not just creating a fun game to play. \tabularnewline \hline
			\raggedright Q3 & ITU & \raggedright Overall, I think the model will help design serious cyber games to meet intended learning objectives and educational effectiveness. I would use it or recommend using it to help design serious cyber games. \tabularnewline \hline
			\end{tabular}%
	\caption{Questionnaire LM-GM versus MOTENS}
	\label{tbl:QuestionnaireLMGMMOTENS}
\end{table}

\subsection{Analysis of the Study}
\begin{figure}[!ht]
\centering
\includegraphics[width=0.9\textwidth]{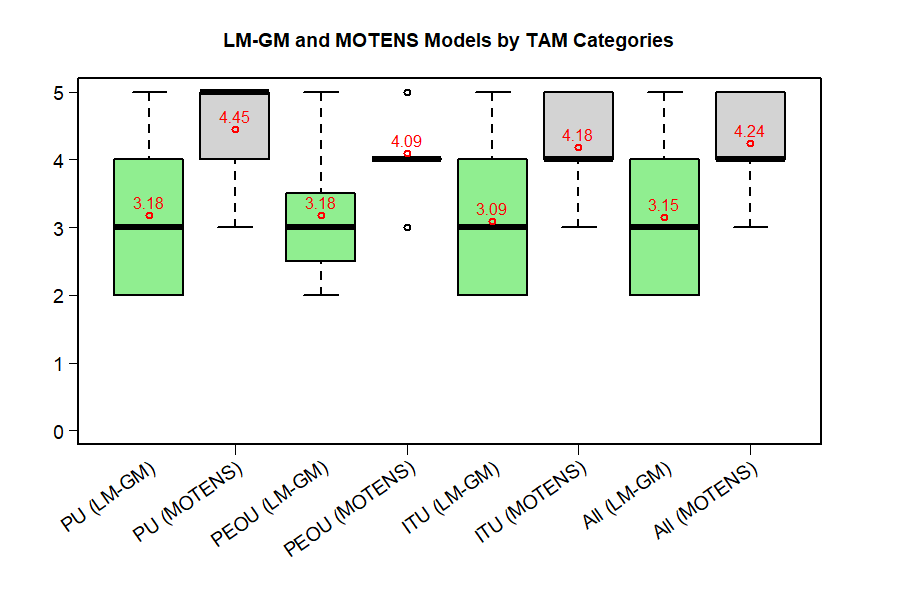}
\caption{LM-GM versus MOTENS by TAM Category}
\label{fig:TAMCategoryLMGMVSMOTENS}
\end{figure}

The results of the analysis to the three questions is shown in \autoref{fig:TAMCategoryLMGMVSMOTENS} with 11 participants. First comparison of TAM scores illustrative case study in \autoref{sec:case_study} and the second is the comparison between MOTENS and LM-GM model.
\paragraph{Comparison to MOTENS in the illustrative case study} The first case study (see \autoref{fig:TAMCategory}) and this study showed comparable results by the TAM category. PU mean as 4.07 and 4.45 and both have IQR 4 to 5. However, the PU median is 4 on illustrative study, compared to PU median 5 testing differences between models. PEOU had similar means of 3.79 and 4.09, with only a difference in IQR of 3 to 4 in the first study compared to IQR 4 in this study. ITU in both studies have the same IQR and similar means of 4.33 and 4.18. The overall perception in the illustrative study of MOTENS excluding generic question 6 was a mean 3.97 compare to the mean of 4.24 in this case study. The overall results of the testing comparative study showed similar results as the illustrative study. 

\paragraph{Comparison of MOTENS and LM-GM Model} The difference between the LM-GM and MOTENS models using the TAM constructs (see \autoref{fig:TAMCategoryLMGMVSMOTENS}) showed MOTENS scoring a consistently higher score than the LM-GM  model.  The IQR of LM-GM for PU and ITU is 2 to 4, whereas MOTENS PU and ITU has IQR of 4 to 5. The LM-GM PEOU IQR is 2.5 to 3.5, whereas MOTENS is 4. The overall perception of LM-GM has a mean of 3.15 with IQR of 2 to 4, compared to MOTENS means of 4.24 and IQR 4 to 5.

\newpage
\section{Summary and Conclusions}
\label{sec:summary} 
This paper proposes a new pedagogical model called MOTENS to design serious cyber games to address current pedagogical models’ limitations in designing and creating serious cyber games. The first illustrative case study (see \autoref{sec:case_study}) the participants were people who would be involved in designing serious cyber games or working in research of cyber security. The results of evaluating the scores were higher for perceived usefulness (PU) and intention to use (ITU) than the perceived ease of use (PEOU) and required a second study to test PEOU. The second case study (see \autoref{sec:case_study2}) targeted students who would use design models to create serious cyber games using a comparative case study between the LM-GM model and the MOTENS model. The PEOU in the second case study improved with IQR of 3 to 4 for MOTENS in the first illustrative case study and IQR 4 in the second comparative case study. The authors note more work needs to be done to improve the PEOU of the MOTENS model. In the creation of the MOTENS model, we identified three key areas in which the MOTENS model has improved on current models: 1) they do not link game mechanics to the learning objectives; 2) high-level model and will not assist in the selection of game mechanics to achieve serious game objectives; 3) are mainly assessed in terms of the quality of their content, not in terms of their intention-based design. We feel the improvements in these three areas using the MOTENS model: 1) Can link the game's mechanics to the target players and select the appropriate game's mechanics to meet the learning objectives, supported by pedagogical learning theory; 2) MOTENS has a five-step process to assist in the game design in stage 4; 3) Learning objectives are built into all six components of the MOTENS model and through the five stages in designing serious games. The case study showed that the participants using the Riskio serious game as an example of using the MOTENS model goes through the five design stages, which will assist you in selecting game mechanics. The first detailed case study also showed how the serious game design can be linked to pedagogical methodology, both the design of the serious game and how this is supported by pedagogical learning theory. The second study a comparison between the LM-GM and our new proposed MOTENS model showed correlated results with similar means, median and IQR. However, the authors suspect that the simple addition of notation for the evaluation against MOTENS might have helped in the participants scoring the LM-GM model. We plan to continue researching the MOTENS model to create serious cyber games for education and verify the MOTENS model educational effectiveness by evaluating the serious games designed using the MOTENS model.

\bibliographystyle{unsrtnat}
\bibliography{references}

\begin{thebibliography}{49}
\providecommand{\natexlab}[1]{#1}
\providecommand{\url}[1]{\texttt{#1}}
\expandafter\ifx\csname urlstyle\endcsname\relax
  \providecommand{\doi}[1]{doi: #1}\else
  \providecommand{\doi}{doi: \begingroup \urlstyle{rm}\Url}\fi

\bibitem[Routledge(2016)]{routledge2016games}
Helen Routledge.
\newblock \emph{Why games are good for business: How to leverage the power of
  serious games, gamification and simulations}.
\newblock Springer, 2016.
\newblock \url{https://link.springer.com/book/10.1057\%2F9781137448989}.

\bibitem[Landers(2014)]{landers2014developing}
Richard~N Landers.
\newblock Developing a theory of gamified learning: Linking serious games and
  gamification of learning.
\newblock \emph{Simulation \& gaming}, 45\penalty0 (6):\penalty0 752--768,
  2014.
\newblock \doi{https://doi.org/10.1177\%2F1046878114563660}.

\bibitem[Hart et~al.(2020)Hart, Margheri, Paci, and Sassone]{hart2020riskio}
Stephen Hart, Andrea Margheri, Federica Paci, and Vladimiro Sassone.
\newblock Riskio: A serious game for cyber security awareness and education.
\newblock \emph{Computers \& Security}, 95:\penalty0 101827, 2020.
\newblock \doi{http://dx.doi.org/10.1016/j.cose.2020.101827}.
\newblock \url{https://www.riskio.co.uk}.

\bibitem[Buil-Gil et~al.(2021)Buil-Gil, Mir{\'o}-Llinares, Moneva, Kemp, and
  D{\'\i}az-Casta{\~n}o]{buil2021cybercrime}
David Buil-Gil, Fernando Mir{\'o}-Llinares, Asier Moneva, Steven Kemp, and
  Nacho D{\'\i}az-Casta{\~n}o.
\newblock Cybercrime and shifts in opportunities during covid-19: a preliminary
  analysis in the uk.
\newblock \emph{European Societies}, 23\penalty0 (sup1):\penalty0 S47--S59,
  2021.
\newblock \url{https://doi.org/10.1080/14616696.2020.1804973}.

\bibitem[Angafor et~al.(2020)Angafor, Yevseyeva, and He]{angafor2020bridging}
Giddeon~N Angafor, Iryna Yevseyeva, and Ying He.
\newblock Bridging the cyber security skills gap: Using tabletop exercises to
  solve the cssg crisis.
\newblock In \emph{Joint International Conference on Serious Games}, pages
  117--131. Springer, 2020.
\newblock \url{https://doi.org/10.1007/978-3-030-61814-8_10}.

\bibitem[Seng~Tan*(2004)]{seng2004students}
Oon Seng~Tan*.
\newblock Students’ experiences in problem-based learning: three blind mice
  episode or educational innovation?
\newblock \emph{Innovations in Education and Teaching International},
  41\penalty0 (2):\penalty0 169--184, 2004.
\newblock \doi{https://doi.org/10.1080/1470329042000208693}.

\bibitem[Greitzer et~al.(2007)Greitzer, Kuchar, and
  Huston]{greitzer2007cognitive}
Frank~L Greitzer, Olga~Anna Kuchar, and Kristy Huston.
\newblock Cognitive science implications for enhancing training effectiveness
  in a serious gaming context.
\newblock \emph{Journal on Educational Resources in Computing (JERIC)},
  7\penalty0 (3):\penalty0 2--es, 2007.
\newblock \doi{https://doi.org/10.1145/1281320.1281322}.

\bibitem[Bloom et~al.(1956)]{bloom1956taxonomy}
Benjamin~S Bloom et~al.
\newblock Taxonomy of educational objectives. vol. 1: Cognitive domain.
\newblock \emph{New York: McKay}, 20:\penalty0 24, 1956.

\bibitem[Hung et~al.(2008)Hung, Jonassen, Liu, et~al.]{hung2008problem}
Woei Hung, David~H Jonassen, Rude Liu, et~al.
\newblock Problem-based learning.
\newblock \emph{Handbook of research on educational communications and
  technology}, 3\penalty0 (1):\penalty0 485--506, 2008.
\newblock
  \url{https://www.routledgehandbooks.com/doi/10.4324/9780203880869.ch38}.

\bibitem[Arnab et~al.(2015)Arnab, Lim, Carvalho, Bellotti, De~Freitas,
  Louchart, Suttie, Berta, and De~Gloria]{arnab2015mapping}
Sylvester Arnab, Theodore Lim, Maira~B Carvalho, Francesco Bellotti, Sara
  De~Freitas, Sandy Louchart, Neil Suttie, Riccardo Berta, and Alessandro
  De~Gloria.
\newblock Mapping learning and game mechanics for serious games analysis.
\newblock \emph{British Journal of Educational Technology}, 46\penalty0
  (2):\penalty0 391--411, 2015.
\newblock \doi{https://doi.org/10.1111/bjet.12113}.

\bibitem[Mitgutsch and Alvarado(2012)]{mitgutsch2012purposeful}
Konstantin Mitgutsch and Narda Alvarado.
\newblock Purposeful by design?: a serious game design assessment framework.
\newblock In \emph{Proceedings of the International Conference on the
  Foundations of Digital Games}, pages 121--128. ACM, 2012.
\newblock \doi{https://doi.org/10.1145/2282338.2282364}.

\bibitem[Amory(2007)]{amory2007game}
Alan Amory.
\newblock Game object model version ii: a theoretical framework for educational
  game development.
\newblock \emph{Educational Technology Research and Development}, 55\penalty0
  (1):\penalty0 51--77, 2007.
\newblock \url{https://link.springer.com/article/10.1007/s11423-006-9001-x}.

\bibitem[Amory and Seagram(2003)]{amory2003educational}
Alan Amory and Robert Seagram.
\newblock Educational game models: conceptualization and evaluation: the
  practice of higher education.
\newblock \emph{South African Journal of Higher Education}, 17\penalty0
  (2):\penalty0 206--217, 2003.
\newblock \url{https://hdl.handle.net/10520/EJC36981}.

\bibitem[Robson et~al.(2015)Robson, Plangger, Kietzmann, McCarthy, and
  Pitt]{robson2015all}
Karen Robson, Kirk Plangger, Jan~H Kietzmann, Ian McCarthy, and Leyland Pitt.
\newblock Is it all a game? understanding the principles of gamification.
\newblock \emph{Business horizons}, 58\penalty0 (4):\penalty0 411--420, 2015.
\newblock \doi{https://doi.org/10.1016/j.bushor.2015.03.006}.

\bibitem[Mullins and Sabherwal(2020)]{mullins2020gamification}
Jeffrey~K Mullins and Rajiv Sabherwal.
\newblock Gamification: A cognitive-emotional view.
\newblock \emph{Journal of Business Research}, 106:\penalty0 304--314, 2020.
\newblock \doi{https://doi.org/10.1016/j.jbusres.2018.09.023}.

\bibitem[Suh et~al.(2018)Suh, Wagner, and Liu]{suh2018enhancing}
Ayoung Suh, Christian Wagner, and Lili Liu.
\newblock Enhancing user engagement through gamification.
\newblock \emph{Journal of Computer Information Systems}, 58\penalty0
  (3):\penalty0 204--213, 2018.
\newblock \doi{https://doi.org/10.1080/08874417.2016.1229143}.

\bibitem[Sansone and Harackiewicz(2000)]{sansone2000intrinsic}
Carol Sansone and Judith~M Harackiewicz.
\newblock \emph{Intrinsic and extrinsic motivation: The search for optimal
  motivation and performance}.
\newblock Elsevier, 2000.
\newblock
  \url{https://www.elsevier.com/books/intrinsic-and-extrinsic-motivation/sansone/978-0-12-619070-0}.

\bibitem[Enzle and Ross(1978)]{enzle1978increasing}
Michael~E Enzle and June~M Ross.
\newblock Increasing and decreasing intrinsic interest with contingent rewards:
  A test of cognitive evaluation theory.
\newblock \emph{Journal of Experimental Social Psychology}, 14\penalty0
  (6):\penalty0 588--597, 1978.
\newblock \doi{https://doi.org/10.1016/0022-1031(78)90052-5}.

\bibitem[Deci and Ryan(2008)]{deci2008self}
Edward~L Deci and Richard~M Ryan.
\newblock Self-determination theory: A macrotheory of human motivation,
  development, and health.
\newblock \emph{Canadian psychology/Psychologie canadienne}, 49\penalty0
  (3):\penalty0 182, 2008.
\newblock \doi{https://doi.org/10.1037/a0012801}.

\bibitem[Malone and Lepper(2005)]{Malone:2005}
Thomas Malone and Mark Lepper.
\newblock Making learning fun: A taxonomy of intrinsic motivations for
  learning.
\newblock \emph{Making Learning Fun: A Taxonomy of Intrinsic Motivations for
  Learning}, 3, 01 2005.

\bibitem[Ros et~al.(2020)Ros, Gonz{\'a}lez, Robles, Tobarra, Caminero, and
  Cano]{ros2020analyzing}
S~Ros, S~Gonz{\'a}lez, A~Robles, LL~Tobarra, A~Caminero, and JESUS Cano.
\newblock Analyzing students’ self-perception of success and learning
  effectiveness using gamification in an online cybersecurity course.
\newblock \emph{IEEE Access}, 2020.
\newblock \doi{http://dx.doi.org/10.1109/ACCESS.2020.2996361}.

\bibitem[B{\'\i}r{\'o}(2014)]{biro2014didactics}
G{\'a}bor~Istv{\'a}n B{\'\i}r{\'o}.
\newblock Didactics 2.0: A pedagogical analysis of gamification theory from a
  comparative perspective with a special view to the components of learning.
\newblock \emph{Procedia-Social and Behavioral Sciences}, 141:\penalty0
  148--151, 2014.
\newblock \doi{https://doi.org/10.1016/j.sbspro.2014.05.027}.

\bibitem[Driscoll(2000)]{driscoll2000psychology}
Marcy~P Driscoll.
\newblock Psychology of learning for instruction.
\newblock \emph{Boston, Allyn and Bacon}, pages 373--396, 2000.

\bibitem[Gagne and Driscoll(1988)]{gagnedriscoll}
Robert~M Gagne and Marcy~P Driscoll.
\newblock \emph{Essentials of learning for instruction}.
\newblock {New Jersey: Prentice Hall, Inc}, 1988.

\bibitem[Unkelos-Shpigel and Hadar(2015)]{unkelos2015gamifying}
Naomi Unkelos-Shpigel and Irit Hadar.
\newblock Gamifying software development environments using cognitive
  principles.
\newblock In \emph{CAiSE Forum}, pages 9--16, 2015.
\newblock \url{http://ceur-ws.org/Vol-1367/paper-02.pdf}.

\bibitem[{National Institute of Standards and Technologies
  (NIST)}()]{NISTcyber}
{National Institute of Standards and Technologies (NIST)}.
\newblock Cyber security framework.
\newblock \url{https://www.nist.gov/cyberframework}.

\bibitem[{NCSE}(2020)]{NCSC:CE}
{NCSE}.
\newblock {National Cyber Security Centre: Cyber Essentials scheme and
  certification}, 2020.
\newblock \url{https://www.ncsc.gov.uk/cyberessentials/overview}.

\bibitem[Rooney(2012)]{rooney2012theoretical}
Pauline Rooney.
\newblock A theoretical framework for serious game design: exploring pedagogy,
  play and fidelity and their implications for the design process.
\newblock \emph{International Journal of Game-Based Learning (IJGBL)},
  2\penalty0 (4):\penalty0 41--60, 2012.
\newblock \url{https://arrow.tudublin.ie/ltcart/28/}.

\bibitem[Gagne and Briggs(1992)]{gagne1992principles}
Robert~M Gagne and Wager~W Briggs, Leslie~J.
\newblock \emph{Principles of instructional design.}
\newblock Harcourt Brace College Publishers, 1992.

\bibitem[Mayer et~al.(2014)Mayer, Bekebrede, Harteveld, Warmelink, Zhou,
  Van~Ruijven, Lo, Kortmann, and Wenzler]{mayer2014research}
Igor Mayer, Geertje Bekebrede, Casper Harteveld, Harald Warmelink, Qiqi Zhou,
  Theo Van~Ruijven, Julia Lo, Rens Kortmann, and Ivo Wenzler.
\newblock The research and evaluation of serious games: Toward a comprehensive
  methodology.
\newblock \emph{British journal of educational technology}, 45\penalty0
  (3):\penalty0 502--527, 2014.
\newblock \doi{https://doi.org/10.1111/bjet.12067}.

\bibitem[McGonigal(2011)]{mcgonigal2011reality}
Jane McGonigal.
\newblock \emph{Reality is broken: Why games make us better and how they can
  change the world}.
\newblock Penguin, 2011.

\bibitem[Fosnot and Perry(1996)]{fosnot:1996}
Catherine~Twomey Fosnot and Randall~Stewart Perry.
\newblock Constuctivism: A psychological theory of learning.
\newblock In \emph{Constructivism: Theory, Perspectives. and Practice},
  chapter~2, pages 9--38. Teachers College Pres, 1996.

\bibitem[Maor(1999)]{maor1999teacher}
Dorit Maor.
\newblock A teacher professional development program on using a constructivist
  multimedia learning environment.
\newblock \emph{Learning Environments Research}, 2\penalty0 (3):\penalty0
  307--330, 1999.
\newblock \url{https://link.springer.com/article/10.1023/A:1009915305353}.

\bibitem[Rolloff(2010)]{rolloff2010constructivist}
Mary Rolloff.
\newblock A constructivist model for teaching evidence-based practice.
\newblock \emph{Nursing Education Perspectives}, 31\penalty0 (5):\penalty0
  290--293, 2010.

\bibitem[Bada and Olusegun(2015)]{bada2015constructivism}
Steve~Olusegun Bada and Steve Olusegun.
\newblock Constructivism learning theory: A paradigm for teaching and learning.
\newblock \emph{Journal of Research \& Method in Education}, 5\penalty0
  (6):\penalty0 66--70, 2015.
\newblock
  \url{https://vulms.vu.edu.pk/Courses/EDU201/Downloads/EDU\%20201\%20(Assignment\%202).pdf}.

\bibitem[Spiro et~al.(2003)Spiro, Collins, Thota, and
  Feltovich]{spiro2003cognitive}
Rand~J Spiro, Brian~P Collins, Jose~Jagadish Thota, and Paul~J Feltovich.
\newblock Cognitive flexibility theory: Hypermedia for complex learning,
  adaptive knowledge application, and experience acceleration.
\newblock \emph{Educational technology}, 43\penalty0 (5):\penalty0 5--10, 2003.
\newblock
  \url{https://www.jstor.org/stable/44429454?seq=1\#metadata_info_tab_contents}.

\bibitem[Perkins(1991)]{perkins1991constructivism}
David~N Perkins.
\newblock What constructivism demands of the learner.
\newblock \emph{Educational technology}, 31\penalty0 (9):\penalty0 19--21,
  1991.
\newblock \url{https://www.jstor.org/stable/44401693}.

\bibitem[Bednar et~al.(1992)Bednar, Cunningham, Duffy, and
  Perry]{bednar1992theory}
Anne~K Bednar, Donald Cunningham, Thomas~M Duffy, and J~David Perry.
\newblock Theory into practice: How do we link.
\newblock \emph{Constructivism and the technology of instruction: A
  conversation}, 8\penalty0 (1):\penalty0 17--34, 1992.
\newblock
  \url{https://www.routledge.com/Constructivism-and-the-Technology-of-Instruction-A-Conversation/Duffy-Jonassen/p/book/9780805812725}.

\bibitem[Dindar et~al.(2021)Dindar, Ren, and
  J{\"a}rvenoja]{dindar2021experimental}
Muhterem Dindar, Lei Ren, and Hanna J{\"a}rvenoja.
\newblock An experimental study on the effects of gamified cooperation and
  competition on english vocabulary learning.
\newblock \emph{British Journal of Educational Technology}, 52\penalty0
  (1):\penalty0 142--159, 2021.
\newblock \doi{https://doi.org/10.1111/bjet.12977}.

\bibitem[Conejo et~al.(2019)Conejo, Gasparini, and
  da~Silva~Hounsell]{conejo2019detailing}
Gabriel~Guebarra Conejo, Isabela Gasparini, and Marcelo da~Silva~Hounsell.
\newblock Detailing motivation in a gamification process.
\newblock In \emph{2019 IEEE 19th International Conference on Advanced Learning
  Technologies (ICALT)}, volume 2161, pages 89--91. IEEE, 2019.
\newblock
  \url{https://slejournal.springeropen.com/articles/10.1186/s40561-019-0106-1}.

\bibitem[Halak(2021)]{halak2021hardware}
Basel Halak.
\newblock \emph{Hardware Supply Chain Security: Threat Modelling, Emerging
  Attacks and Countermeasures}.
\newblock Springer Nature, 2021.
\newblock \url{https://www.springer.com/gp/book/9783030627065}.

\bibitem[{NCSC}(2021)]{NCSC:10}
{NCSC}.
\newblock {National Cyber Security Centre: 10 Step to Cyber Security: Guidance
  on how organisations can protect themselves in cyberspace}, 2021.
\newblock \url{https://www.ncsc.gov.uk/collection/10-steps-to-cyber-security}.

\bibitem[Seng(2000)]{seng2000reflecting}
TO~Seng.
\newblock Reflecting on innovating the academic architecture for the 21st
  century.
\newblock \emph{Educational Developments}, 1:\penalty0 8--10, 2000.

\bibitem[Yusoff(2010)]{yusoff2010conceptual}
Amri Yusoff.
\newblock \emph{A conceptual framework for serious games and its validation}.
\newblock PhD thesis, University of Southampton, 2010.
\newblock \url{https://eprints.soton.ac.uk/171663/}.

\bibitem[Nguyen and Pham(2020)]{nguyen2020design}
Tuan~Anh Nguyen and Hiep Pham.
\newblock A design theory-based gamification approach for information security
  training.
\newblock In \emph{2020 RIVF International Conference on Computing and
  Communication Technologies (RIVF)}, pages 1--4. IEEE, 2020.
\newblock \doi{https://doi.org/10.1109/RIVF48685.2020.9140730}.

\bibitem[Khan et~al.(2011)Khan, Alghathbar, Nabi, and
  Khan]{khan2011effectiveness}
Bilal Khan, Khaled~S Alghathbar, Syed~Irfan Nabi, and Muhammad~Khurram Khan.
\newblock Effectiveness of information security awareness methods based on
  psychological theories.
\newblock \emph{African Journal of Business Management}, 5\penalty0
  (26):\penalty0 10862--10868, 2011.
\newblock \url{https://api.semanticscholar.org/CorpusID:154298999}.

\bibitem[Davis(1989)]{Davis:1989}
Fred~D Davis.
\newblock Perceived usefulness, perceived ease of use, and user acceptance of
  information technology.
\newblock \emph{MIS Quarterly}, pages 319--340, 1989.
\newblock \doi{https://doi.org/10.2307/249008}.

\bibitem[Yusoff et~al.(2010)Yusoff, Crowder, and Gilbert]{yusoff2010validation}
Amri Yusoff, Richard Crowder, and Lester Gilbert.
\newblock Validation of serious games attributes using the technology
  acceptance model.
\newblock In \emph{2010 Second International Conference on Games and Virtual
  Worlds for Serious Applications}, pages 45--51. IEEE, 2010.
\newblock \doi{https://doi.org/10.1109/VS-GAMES.2010.7}.

\bibitem[Wohlin et~al.(2012)Wohlin, Runeson, H{\"o}st, Ohlsson, Regnell, and
  Wessl{\'e}n]{Wohlin:2012}
Claes Wohlin, Per Runeson, Martin H{\"o}st, Magnus~C Ohlsson, Bj{\"o}rn
  Regnell, and Anders Wessl{\'e}n.
\newblock \emph{Experimentation in software engineering}.
\newblock Springer Science \& Business Media, 2012.
\newblock ISBN 3642290434.
\newblock \url{https://www.springer.com/gp/book/9783642290435}.

\end{thebibliography}






\end{document}